\documentclass[12pt]{iopart}

\usepackage{iopams}  

\usepackage{graphicx}
\usepackage{amssymb}
\usepackage{epsfig}
\usepackage{dcolumn}% Align table columns on decimal point
\usepackage{bm}% bold math
\usepackage{bbm}
\usepackage{dsfont}
\usepackage{color}
\usepackage[up]{subfigure}
\usepackage{cite}

\begin{document}

\title{Geometry of the generalized Bloch sphere for qutrits}

 \author{Sandeep K. Goyal$^{1*}$, B. Neethi Simon$^{2}$, Rajeev Singh$^{3\dagger}$ \& Sudhavathani Simon$^{4}$}

\address{$^1$Institute for Quantum Science and Technology,
University of Calgary, Alberta T2N~1N4, Canada\\
$^2$Department of Mechanical Engineering,
SSN College of Engineering,
 SSN Nagar, OMR, Chennai 603~110, India\\
$^3$The Institute of Mathematical Sciences, Tharamani, Chennai 600~113\\
$^4$Department of Computer Science,
Women's Christian College,  Chennai 600~006, India}
\ead{$^*$sandeep.goyal@ucalgary.ca,~$^\dagger$rajeev@imsc.res.in}
\vspace{10pt}
\begin{indented}
\item \today
\end{indented}

\begin{abstract}
The geometry of the generalized Bloch sphere $\Omega_3$, the state
space of a qutrit, is studied. Closed form expressions for $\Omega_3$,
its boundary $\partial \Omega_3$, and the set of extremals
$\Omega_3^{\rm ext}$ are obtained by use of an elementary observation.
These expressions and analytic methods are used to classify the 28
two-sections and the 56 three-sections of $\Omega_3$ into unitary
equivalence classes, completing the works of earlier authors.  It is
shown, in particular, that there are families of two-sections and of
three-sections which are equivalent geometrically but not unitarily, a
feature that does not appear to have  been appreciated earlier. A
family of three-sections of obese-tetrahedral shape whose symmetry
corresponds to the 24-element tetrahedral point group $T_d$ is
examined in detail. This symmetry is traced to the natural reduction of
the adjoint representation of $SU(3)$, the symmetry underlying
$\Omega_3$, into direct sum of the two-dimensional and the two
(inequivalent) three-dimensional irreducible representations of $T_d$.
\end{abstract}

\maketitle

\section{Introduction}
 
States of a $d$-level quantum system are in one-to-one correspondence
with positive semidefinite unit-trace operators on a $d$-dimensional Hilbert space
$\mathcal{H}$. The defining properties of these density operators or density 
matrices $\rho$ are\,:   (i)~$\rho^{\dagger} = \rho,$
  (ii)~$\rho \ge 0,$ and (iii)~${\rm tr}\,\rho = 1$. 
The collection of all possible density operators 
of a $d$-level system is its state space $\Omega_d$. 
 Pure states correspond to the further matrix condition  
$\rho^2= \rho$ which is equivalent to the scalar condition 
${\rm tr}\,\rho^2 =1$.  All other states are mixed states. 
It follows from definition  that 
{\em   the quantum state space is a convex set}. That is,  
a convex combination of density operators 
always results in density operator\,: $\rho_j \in \Omega_d$,  
$j=0,\,1,\,\cdots,\,\ell$ implies $\sum_j p_j\rho_j \in \Omega_d$ 
for every set $\{\,p_j\,\}$ with $p_j >0,\;\, \sum_j p_j=1$. 

The importance of a good understanding of the geometry of $\Omega_d$ is 
evident. 
 The structure of $\Omega_2$, the state space of a two-level system or 
qubit is  simple\,: it is the Bloch (or Poincar\'{e}) sphere. 
But the structure of $\Omega_d$, the {\em generalized Bloch sphere}, is 
much richer for $d\ge 3$,  and has attracted the attention 
of many authors\,\cite{Bengtsson2006,bloore,murthy1,murthy2,hioe,pottinger85,kimura,Kimura2005,mendas,jakobczyk,kryszewski,byrd,IvanovNishi10,Ivanov10,Kolenderski10,Mendas2008,bertlmann08,krammer09,bertlmann08a,bertlmann08b,bertlmann09,Kossakowski2003,Kimura2004,Byrd2011,Byrd2010,Wodkiewicz2009,Dixit2008,Simon2009,dietz,Boya2008,schirmer04,bourdon04,Appleby2011,Vinjanampathy2009,Tilma2002,Mosseri2001,Tay2008,Mendas2009,Taguchi2009,kurzyński2009,SolisProsser11,Bengtsson2013,Tabia2013,Sarbicki2013,Kus2001,Verstraete2002,Singh2013,Sabapathy2013}.

In order to get a quick partial insight into this richness several authors 
 have looked at sections of $\Omega_d$. The original motivation 
for considering such sections comes from Bloore\,\cite{bloore} who 
 examined the five-section of $\Omega_3$ corresponding to 
density matrices whose entries are all {\em real}, and noted that this 
five-section has a three-section of the shape of a tetrapack. 
 (As we shall see below, there exist also other three-sections of 
$\Omega_3$ of precisely this shape.)  
Two-sections have been considered by Kimura\,\cite{kimura}, Kimura and 
Kossakowski\,\cite{Kossakowski2003}, Menda\v{s}\,\cite{mendas}, 
Jak\'obczyk and Siennicki\,\cite{jakobczyk}, Kryszewski and 
Zachcial\,\cite{kryszewski}, and Sarbicki and Bengtsson~\cite{Sarbicki2013}. 
Similarity between state space and multi-Higgs doublets and Majorana
representation for state space has been studied~\cite{IvanovNishi10,
Ivanov10, Kolenderski10}.
 The work of Mendas\,\cite{mendas} 
aims also at a classification of the 56 three-sections of  
 $\Omega_3$  using Monte Carlo sampling. 

We present in this paper an analysis of the structure of $\Omega_3$  
 using analytic methods. For purpose of comparison we begin in Section~2 
with a brief consideration of $\Omega_2$. In Section~3 we use the elementary 
observation that there exists within the `unit sphere' $S^7$ in $\mathds{R}^8$  no 
singular  (hermitian) matrix which is indefinite
%(i.e., it has both positive and negative eigenvalues)
to derive {\em closed-form 
expressions} for $\Omega_3$, its boundary $\partial \Omega_3$, and for the set of 
extremals $\Omega_3^{\rm ext}$. We use these closed-form expressions to 
classify in Section~4 the 28 two-sections of $\Omega_3$ into five unitary 
equivalence 
classes of four geometrically different shapes. 
The 56 three-sections are shown in Section~5 to group themselves into ten unitary 
inequivalence classes of only seven geometrically different shapes, and the 
results are 
contrasted with those of Mendas\,\cite{mendas}. 

One family of three-sections has an interesting geometric shape, namely, the obese-tetrahedron, 
 having symmetry that corresponds to the 24-element  tetrahedral group $T_d$ familiar 
from 
the context of 
point groups. We study this family in considerable detail in Section~6, and trace this  
$T_d$ symmetry consisting of `proper and improper' rotations to a discrete subgroup of 
 proper rotations in $SO(3) \subset SU(3)$. This is the subgroup 
for 
which the 
eight-dimensional adjoint representation of $SU(3)$ reduces to direct sum of 
one two-dimensional and two (inequivalent) three-dimensional irreducible 
representations of $T_d$. We conclude in Section~7 with some final remarks.

\section{Two-level systems: Bloch sphere}\label{se}
The density operator in this case is a hermitian, positive
(semidefinite) $2\times 2$
matrix of unit trace. The unit matrix $\sigma_0 \equiv \mathds{1}_{2\times 2}$   
along with the Pauli matrices $\sigma_1,\,\sigma_2,\,\sigma_3$ satisfy 
\begin{eqnarray}
 \sigma_j\sigma_k + \sigma_k\sigma_j &&= 2\sigma_0\delta_{jk},\nonumber\\
 \sigma_j\sigma_k - \sigma_k\sigma_j &&= 2i\epsilon_{jk\ell}\sigma_\ell,\nonumber\\
{\rm tr}\,\sigma_{\mu}\sigma_{\nu} &&= 2\delta_{\mu\,\nu},\quad\mu,~\nu=
0,\,1,\,2,\,3;\;\;j,k,\ell = 1,\,2,\,3.\nonumber
\end{eqnarray}
They form a complete set of hermitian orthogonal matrices,
 and since any $2\times 2$ hermitian matrix can be expressed as a unique linear
combination of these matrices, with {\em real} coefficients, we may write 
\begin{equation}
\rho = \rho({\bm n}) = \frac{1}{2}\left(\sigma_0 +
  {\bm n}\cdot{\bm \sigma}\right),
\end{equation}
for some vector ${\bm n}= (n_1,n_2,n_3) \in\mathds{R}^3$.
That the coefficient of $\sigma_0$ is $1/2$ follows from and ensures
${\rm tr}\rho = 1$. Positivity of this hermitian unit trace expression
 demands $|{\bm n}|^2 = {\bm n}\cdot\mbox{\boldmath $n$} \le 1$. 
Indeed, the eigenvalues
of $\rho({\bm n})$ are
seen to be $(1\pm|{\bm n}|)/2$.

It follows that the states are in one-to-one correspondence with the
 points on or inside the closed unit ball in $\mathds{R}^3$, centered 
at the origin of $\mathds{R}^3$. This is
the Bloch or Poincar\'{e} ball. Points on the boundary $S^2$
correspond to pure states: the unit vector $|\psi\rangle =
e^{i\alpha}\left(\begin{array}{c}\cos\theta/2\\
    e^{i\phi}\sin\theta/2\end{array}\right)$  which corresponds to the
one-dimensional projector
\begin{eqnarray}
\rho(\theta,~\phi) &=& \frac{1}{2}\left(
\begin{array}{cc}
1+\cos\theta & \sin\theta e^{-i\phi}\\
\sin\theta e^{i\phi} & 1-\cos\theta
\end{array}\right),
\end{eqnarray}
is seen to be represented  by unit vector
$n(\theta,~\phi)~=~(\sin\theta\cos\phi,~\sin\theta\sin\phi,~\cos\theta)
\in S^2$. It is clear that
orthogonal pure states occupy antipodal points on $S^2$. 
The interior points of the unit ball correspond to mixed states. In
particular the centre ${\bm n} = 0$ corresponds to the maximally mixed state
$\rho = \frac{1}{2}\sigma_0$.

 The special property of this convex set or state space $\Omega_2 $ 
which is not
shared by $n$-level systems with $n>2$ is this: every boundary point is
an extremal. 
 There exists another, related aspect in which the quantum two-level
system differs from $n$-level systems with $n\ge 3$. Unitary
evolutions $\rho \to \rho' = U\rho U^{\dagger},~U\in SU(2)$ act on $\Omega_2 
\subset
\mathds{R}^3$ through $SO(3)$ rotations arising from the adjoint
representation of $SU(2)$. And $S^2$, the boundary of $\Omega_2$ is a
 {\em single orbit} under this action. 
 This does not hold for $\Omega_d, \; d\ge 3$.  While it is true that 
 unitary evolutions $\rho\to\rho' = U\rho U^{\dagger},~U\in SU(d)$ 
 continue to act as 
 rotations on $\Omega_d \subset \mathds{R}^{d^2 -1}$ through the adjoint
representation, the $SU(d)$ orbit of pure states which necessarily corresponds 
to the $2(d-1)$-dimensional manifold $SU(d)/SU(d-1) = CP^{d-1}$ is a measure zero 
subset of the $(d^2-2)$-dimensional boundary $\partial \Omega_d$\,: 
 the boundary of $\Omega_d$ consists of {\em a continuum of
  distinct $SU(d)$ orbits}.

\section{Bloch representation for the qutrit}
In place of 
the three Pauli matrices we now need the following  eight Gell-Mann
$\lambda$-matrices to describe a generalization of the  Bloch ball
representation of qubit  to the case of {\em three-level system} or
{\em qutrit}: 
%\begin{center}
\begin{eqnarray*}
\lambda_1 =
\left( \begin{array}{ccc}
0&1&0\\
1&0&0\\
0&0&0\\
\end{array} \right),
\lambda_2=
\left( \begin{array}{ccc}
0&-i&0\\
i&0&0\\
0&0&0\\
\end{array} \right),
\lambda_3 =
\left( \begin{array}{ccc}
1&0&0\\
0&-1&0\\
0&0&0\\
\end{array} \right),\nonumber\\
\lambda_4  =
\left( \begin{array}{ccc}
0&0&1\\
0&0&0\\
1&0&0\\
\end{array} \right),
\lambda_5 =
\left( \begin{array}{ccc}
0&0&-i\\
0&0&0\\
i&0&0\\
\end{array} \right),
\lambda_6=
\left( \begin{array}{ccc}
0&0&0\\
0&0&1\\
0&1&0\\
\end{array} \right),\nonumber\\
\lambda_7 =
\left( \begin{array}{ccc}
0&0&0\\
0&0&-i\\
0&i&0\\
\end{array} \right),
\lambda_8=
\frac{1}{\sqrt{3}}\left( \begin{array}{ccc}
1&0&0\\
0&1&0\\
0&0&-2\\
\end{array} \right).\nonumber
\end{eqnarray*}
%\end{center}
These matrices are familiar as generators of the unimodular unitary group $SU(3)$ 
in its 
defining representation. Just like the Pauli matrices these form a {\em
  complete} set of hermitian, traceless, trace-orthogonal matrices:
\begin{equation}
{\rm tr}\lambda_k\lambda_l = 2\delta_{kl},~~k,l = 1,~2,~\cdots,~8.
\end{equation}
The algebraic structure of these matrices is determined by the product
property\,\cite{Khanna,Mallesh1997,Arvind1997,Ercolessi2001}
\begin{equation}
\lambda_j\lambda_k =  \frac{2}{3}\delta_{jk} + d_{jkl}\lambda_l
+if_{jkl}\lambda_l. 
\end{equation}
Separating this product into hermitian and antihermitian parts, we have
\begin{eqnarray}
\lambda_j\lambda_k -\lambda_k\lambda_j &=& 2if_{jkl}\lambda_l,\nonumber\\
\lambda_j\lambda_k +\lambda_k\lambda_j &=& \frac{4}{3}\delta_{jk} +
2d_{jkl}\lambda_l.
\end{eqnarray}
The expansion coefficients $f_{jkl}$, the structure constants of the Lie
algebra of $SU(3)$, are {\em totally antisymmetric} in their indices,
whereas $d_{jkl}$ are {\em totally symmetric}. It is useful to list the
numerical values of all the independent nonvanishing components of
$f_{jkl},\,d_{jkl}$\,\cite{Khanna,Mallesh1997,Arvind1997,Ercolessi2001}:
\begin{equation}
\eqalign{
f_{123} = 1,~f_{458} = f_{678}= \frac{\sqrt{3}}{2},\\
f_{147} = f_{246} = f_{257} = f_{345} = f_{516} = f_{637} =
\frac{1}{2};\\
d_{118} =d_{228} =d_{338} = -d_{888} = \frac{1}{\sqrt{3}},\\
d_{146} =d_{157}= -d_{247}=d_{256}=\frac{1}{2},\\
d_{344} = d_{355} = -d_{366}= -d_{377}=\frac{1}{2},\\
d_{448} = d_{558} =d_{668} =d_{778} = -\frac{1}{2\sqrt{3}}.}
\end{equation}
These $\lambda$-matrices can be used to describe any $3\times 3$
density matrix $\rho$ in terms of a corresponding 8-dimensional 
{\em real} vector ${\bm n}$:
\begin{equation}
\rho({\bm n}) = \frac{1}{3}\left( \mathds{1} + \sqrt{3}\,{\bm n}.{\bm
    \lambda}\right),~{\bm n}\in \mathds{R}^8. \label{rho} 
\end{equation}
Hermiticity of $\rho({\bm n})$ is ensured by that of the $\lambda$-matrices and
reality of ${\bm n}$, and unit trace property is ensured by the prefactor
$\frac{1}{3}$. Thus, it remains only to choose 
(restrict) ${\bm n}\in \mathds{R}^8$
 such that the matrix $\mathds{1} + \sqrt{3}\,{\bm n}\cdot{\bm 
\lambda}$ is positive semidefinite.

Let $\Omega_3$ be the set of all points ${\bm n}\in \mathds{R}^8$ such that
$\mathds{1} + \sqrt{3}\,{\bm n}\cdot{\bm \lambda} \ge 0$. Then, by definition,  
$\Omega_3\subset
\mathds{R}^8$ is our state space or {\em generalized Bloch sphere}. We know 
that $\Omega_3$ is a closed 
convex
set.  Our aim here is to develop an understanding of the structure of 
$\Omega_3$, and we
begin  by giving {\em closed form expressions} for both its boundary 
$\partial \Omega_3$ and the set of all extremals $\Omega_3^{\rm ext}$, leading 
eventually to a closed form  expression for the convex $\Omega_3$ itself.

\subsection{Extremal points of $\Omega_3$}

We begin  our analysis by defining a {\em star-product} 
$\mathds{R}^8\to\mathds{R}^8$ on vectors ${\bm 
n}\in \mathds{R}^8$ through~\cite{Khanna,Mallesh1997,Arvind1997,Ercolessi2001} 
\begin{equation}
({\bm n}*{\bm n})_l = \sqrt{3}\,d_{jkl}\,n_jn_k.
\end{equation}
As an immediate consequence and application we have the following 
expression for the
square of any  traceless hermitian matrix ${\bm n}\cdot{\bm  \lambda}$:
\begin{equation}
({\bm n}\cdot{\bm  \lambda})^2 = {\bm n}\cdot{\bm  \lambda}\,{\bm n}\cdot{\bm \lambda} = 
\frac{2}{3}{\bm n} \cdot{\bm n} + \frac{1}{\sqrt{3}}\,{\bm n}*{\bm 
n}\cdot{\bm \lambda}.
\label{eqn9}
\end{equation}

Now the necessary and sufficient condition for a density operator 
$\rho({\bm n})$ 
to correspond to a pure state, and hence to an extremal of $\Omega_3$, is
that it be a one dimensional projector, i.e, 
$\rho({\bm n})^2 = \rho({\bm n})$.  So we compute
\begin{equation}
\rho({\bm n})^2 = \frac{1}{9} \left( \mathds{1} +2{\bm n}\cdot{\bm n} + 
2\sqrt{3}\,{\bm n}\cdot{\bm \lambda} +
  \sqrt{3}\,{\bm n}*{\bm n}\cdot{\bm \lambda}\right). 
\end{equation}
Equating this expression to $\rho({\bm n}) = (\mathds{1} + \sqrt{3}\,{\bm n}.{\bm
    \lambda})/3$, Eq.\,(\ref{rho}),  
%%%%%%%%%%%
we obtain a pair of 
 constraints  
\begin{equation}
{\bm n}\cdot{\bm n} = 1, ~~~{\bm n}*{\bm n} = {\bm n}
\end{equation}
as the defining conditions on points ${\bm n}\in \mathds{R}^8$ in
order that $\rho({\bm n}) \in \Omega^{\rm ext}_3 \subset S^7$, the set of all 
the extremals of our state space $\Omega_3$.  Since we know that these 
extremals constitute also 
$CP^2$, the manifold of pure states of a three-level quantum system,
we may write: 
\begin{eqnarray}
\Omega^{\rm ext}_3 = 
CP^2 &=& \left\{ {\bm n}\in\mathds{R}^8\,|\, ~{\bm n}\cdot{\bm n} = 
1,~{\bm
    n}*{\bm n}={\bm n}\right\}. \label{extremal}
\end{eqnarray}

Having thus described in {\em closed form } the four-parameter family of extremals
of $\Omega_3$, we now move on to develop an expression for the boundary
$\partial \Omega_3$.

\subsection{ Closed-form expressions for the Bloch sphere $\Omega_3$ and its 
boundary $\partial\Omega_3$}

It is clear that the boundary of $\Omega_3$ comprises density matrices
which are singular. Thus the boundary points ${\bm n}\in\mathds{R}^8$ should
necessarily satisfy the condition
\begin{eqnarray}
&& \det\rho({\bm n}) = 0 = \det(\mathds{1} + \sqrt{3}\,{\bm n}\cdot{\bm 
\lambda)},\nonumber\\
&& {\bm n}\cdot{\bm \lambda} =\left(\begin{array}{ccc}
n_3+\frac{1}{\sqrt{3}}n_8&n_1-in_2&n_4-in_5\\
n_1+in_2&-n_3+\frac{1}{\sqrt{3}}n_8&n_6-in_7\\
n_4+in_5&n_6+in_7&-\frac{2}{\sqrt{3}}n_8\\
\end{array}\right). 
\end{eqnarray}
Writing out this requirement in detail, we have
\begin{eqnarray}
3(n_1^2+n_2^2+n_3^2+n_4^2+n_5^2+n_6^2+n_7^2+n_8^2)&&\nonumber\\ 
~~- 6n_8\left(n_1^2+n_2^2+n_3^2-\frac{n_4^2+n_5^2+n_6^2+n_7^2}{2}
- \frac{n_8^2}{3}\right)&&\nonumber\\ 
~~~~~~-6\sqrt{3}(n_1n_4n_6+n_1n_5n_7+n_2n_5n_6-n_2n_4n_7)&&\nonumber\\
~~~~~~~~~~-3\sqrt{3}n_3(n_4^2+n_5^2-n_6^2-n_7^2) &&=1.  \label{***}
\end{eqnarray}
Every boundary point ${\bm n}\in \partial\Omega_3$ should necessarily
satisfy this {\em cubic} constraint.  
 It is readily verified that $\mathds{1}+\sqrt{3}\,{\bm n}\cdot{\bm 
\lambda}$ is
{\em necessarily indefinite}  if ${\bm n}\cdot{\bm n} >1$, 
 and therefore nonnegativity of 
$\rho$ demands that 
${\bm n}$ should  satisfy the additional inequality
\begin{equation}
|{\bm n}|^2 \equiv \sum_{j=1}^{8}n_j^2 \le 1.\label{****}
\end{equation}
As the final step, we will make use of the following elementary observation.

\vskip 0.2cm
\noindent
{\bf Theorem}: No {\em singular matrix} of the form
 $\mathds{1} + \sqrt{3}\,{\bm n}\cdot{\bm \lambda}$, {\em with} $|{\bm 
n}|\le 1$, can be indefinite. 

\noindent
{\em Proof}: Singularity of $\mathds{1} +
\sqrt{3}\,{\bm n}\cdot{\bm \lambda}$ implies that the matrix
 ${\bm n}\cdot{\bm \lambda}$ should necessarily have 
$-\frac{1}{\sqrt{3}}$ as one of
 its eigenvalues. Tracelessness of ${\bm n}\cdot{\bm \lambda}$ then 
implies that the
other two eigenvalues be of the form $\mu, \frac{1}{\sqrt{3}} -\mu$. But
the condition $|{\bm n}|\le 1$ is equivalent to the requirement that the sum
of  squares of the eigenvalues of ${\bm n}\cdot{\bm \lambda}$ be 
bounded by $2$ as may be seen from Eq.~(\ref{eqn9}), i.e., 
$\left(\frac{1}{\sqrt{3}}\right)^2 + \mu^2 + \left(\frac{1}{\sqrt{3}}
  -\mu\right)^2 \le 2$. The last inequality obviously forces $\mu$ to
the range 
\begin{eqnarray}
-\frac{1}{\sqrt{3}}\le \mu \le \frac{2}{\sqrt{3}},\label{***1} 
\end{eqnarray}
proving
our assertion that {\em every singular matrix} 
  $\mathds{1}+\sqrt{3}{\bm n}\cdot{\bm \lambda}$ {\em with} $|{\bm n}| 
\le 
1$ is nonnegative.

Thus  the cubic equation  (\ref{***}) and the quadratic restriction 
 (\ref{****}) together
present  a complete characterization of the `Bloch sphere'
$\partial \Omega_3$: {\em the boundary of $\Omega_3$ is that portion 
of the
cubic surface (Eq.\ref{***}) contained within the unit ball $|{\bm 
n}| 
\le
1$ in $\mathds{R}^8$}. That $\partial \Omega_3$ is a closed surface
 was already clear 
from the fact that it is the boundary of a closed convex set in
finite dimension.

One might have noticed that the
coefficients of the cubic terms in (\ref{***}) exactly match the
numerical  values of the symmetric $d_{jkl}$ symbols. Indeed, the 
quadratic and
cubic terms in  (\ref{***}) equal respectively $3{\bm n}\cdot{\bm n}$ and
$-2{\bm n}*{\bm n}\cdot{\bm n}$. We may thus characterize the Bloch sphere
$\partial \Omega_3$ 
in a form which is as elegant as the characterization of $CP^2$ in
Eq.\,(12): 
\begin{equation}
\partial\Omega_3 = \left\{{\bm n}\in\mathds{R}^8\,\,|\,\, 3{\bm 
n}\cdot{\bm n}
  -2{\bm n}*{\bm n}\cdot{\bm n} = 1,~~{\bm n}\cdot{\bm n} \le 
1\right\}. \label{surface}
\end{equation}
To go hand in hand with (\ref{extremal}) and (\ref{surface}), the state space 
$\Omega_3$  
 itself may be fully characterized in the following manner:
\begin{equation}
\Omega_3 = \left\{{\bm n}\in\mathds{R}^8\,\,|\,\, 3{\bm 
n}\cdot{\bm n}
  -2{\bm n}*{\bm n}\cdot{\bm n} \le 1,~~{\bm n}\cdot{\bm n} \le 
1\right\}. \label{solid}
\end{equation}
At the risk of repetition we stress that $\Omega_3$ is a {\em portion}
of the solid sphere in $\mathds{R}^8$. It is not $SO(8)$ invariant,
but invariant under a 8-parameter subgroup thereof.

We have thus characterized $\Omega_3$ and $\partial\Omega_3$ as
compactly as $\Omega_3^{\rm ext}$ in~(\ref{extremal}): $\partial
\Omega_3$ corresponds to {\em saturation of one inequality}, the first
one, in~(\ref{solid}) while $\Omega_3^{\rm ext} \subset \partial \Omega_3$
corresponds to {\em saturation of both}.
That is, $\Omega_3^{\rm ext}$ is that subset of $\partial \Omega_3$ for
which ${\bm n}\cdot{\bm n} = 1$.
 These are the only two possibilities for saturation of the inequalities 
in (\ref{solid}), for the positivity condition  $\rho({\bm n}) \ge 0$ implies  
that under saturation of the second inequality the first one is automatically 
saturated. 
{\em This coordinate-free characterization is not only 
compact, but also 
 renders $SU(3)$ invariance of $\Omega_3, \,\partial\Omega_3$, and 
$\Omega_3^{\rm ext}$ manifest}, and is strictly analogous to 
\begin{eqnarray*}
\Omega_2&=& B_3 = \left\{{\bm n}\in\mathds{R}^3\,\,|\,\, {\bm 
n}\cdot{\bm n} \le 1\right\}, \nonumber\\
\Omega_2^{\rm ext}= \partial \Omega_2 &=& S^2 = CP^1= \left\{{\bm 
n}\in\mathds{R}^3\,\,|\,\, 
{\bm 
n}\cdot{\bm n} = 1\right\},  
\end{eqnarray*}
of the simpler case of two-dimensional Hilbert space.

{\bf Remark}. Our derivation of closed-form expressions for $\Omega_3$ and its boundary 
 $\partial \Omega_3$ is based on the elementary observation that 
a singular matrix of the form $\mathds{1}~+~\sqrt{3}{\bm n}\cdot~{\bm \lambda},\;\,{\bm 
n}\in \mathds{R}^8$ cannot be indefinite if ${\bm n}\cdot{\bm n}\le 1$. 
 We could have used instead the following general 
procedure\,\cite{Simon1987,Simon1987b}. 
Suppose we wish to characterize
positivity of the $d\times d$ matrix $\mathds{1}~+~H$ where $H$ is hermitian. 
 We may begin by 
evaluating $\det (y~\mathds{1}~+~H~)$ where $y$ is a real variable. The result 
will clearly be a polynomial in $y$ of degree $d$\,: 
\begin{eqnarray*}
P(y)&\equiv& \det (y~\mathds{1} + H )\nonumber\\
   &=& y^d +c_1y^{d-1} + c_2y^{d-2} + \;\cdots\;+ c_{d-1}y + c_d.
\end{eqnarray*}
All the coefficients $c_j$ are determined by the (unitarily invariant) traces 
${\rm tr}\,H,\,{\rm tr}\,H^2,\,\cdots \,,\,{\rm tr}\,H^d$ of $H$. In particular, 
$c_1 = {\rm tr}\,H$ and $c_d = \det H$. ($\,\det H$ too is determined by the 
 above invariant traces.) It is now clear that a necessary and sufficient set of conditions 
for positivity of $\mathds{1}~+~H$ is 
\begin{eqnarray*}
P(y)\,\Big|_{y=1} \ge 0,\;\,\frac{ d P(y)}{dy}\,\Big|_{y=1} \ge
0,\;\cdots\,,\; \frac{d^{d-1} P(y)}{dy^{d-1}}\,\Big|_{y=1} \ge 0\,.
\end{eqnarray*}
This characterization of positivity was used in \cite{Simon1987,Simon1987b} 
to characterize variance matrix of a multimode quantum system
  in a covariant manner. More recently, similar  
procedure has been used to characterize $\Omega_d$\,\cite{byrd}. Since we are dealing 
with the particular case $d=3$,  
 rather than  arbitrary $d$, we have preferred to base our characterization 
of $\Omega_3$ on the elementary observation mentioned above, rather than on this 
general procedure.

\subsection{Special spheres associated with $\Omega_3$}
 As noted earlier, the boundary $\partial \Omega_3$ corresponds to singularity of 
 the matrix $\mathds{1} +\sqrt{3}\,{\bm n}\cdot{\bm \lambda}$,  implying that the 
 eigenvalues of the traceless matrix ${\bm n}\cdot{\bm \lambda}$ 
should necessarily be of the form  
\begin{eqnarray}
-\frac{1}{\sqrt{3}},\;\;\mu,\;\; \frac{1}{\sqrt{3}}-\mu, \label{***2}
\end{eqnarray}
 and nonnegativity of $\mathds{1} +\sqrt{3}\,{\bm n}\cdot{\bm \lambda}$ 
is seen to restrict $\mu$ to the range given in (\ref{***1}), obtained earlier as
consequence of $|{\bm n}|\le 1$.  Norm of ${\bm n}\in \mathds{R}^8$ is determined 
 by the eigenvalues of ${\bm n}\cdot{\bm \lambda}$, and we have 
\begin{eqnarray}
|{\bm n}| &\equiv& \sqrt{{\bm n}\cdot{\bm n}} =
\sqrt{\frac{1}{2}\,{\rm tr}\,({\bm n}\cdot{\bm \lambda})^2}\nonumber\\
&=& \sqrt{\frac{1}{2} \left.\left[\left(\frac{1}{\sqrt{3}}\right)^2 +\mu^2 +\left( 
\frac{1}{\sqrt{3}}-\mu\right)^2\right]\right.\,}, \label{***3}
\end{eqnarray}
whose minimum value is 1/2 and corresponds to $\mu = (2\sqrt{3})^{-1}$. 
 This value of $\mu$ is the `midpoint' of the values 
 $\mu = -\frac{1}{\sqrt{3}},\, \frac{2}{\sqrt{3}}$ at which the maximum norm, 
$|{\bm n}|= 1$, obtains. 

Thus the boundary $\partial \Omega_3$ can never stray into the interior of 
the eight-dimensional solid sphere of radius 1/2 contained in $\Omega_3$. 
Since the maximum possible norm for ${\bm n}$ is unity the boundary $\partial 
\Omega_3$, and hence the state space $\Omega_3$ itself, can never stray into the 
exterior of the unit ball in $\mathds{R}^8$; the latter result is a fact we knew all 
along. 

Now, these two concentric balls in $\mathds{R}^8$ of radii 1 and 1/2 respectively are 
dual to one another in the following sense.   
  The eigenvalues of ${\bm n}\cdot{\bm \lambda}$ corresponding to 
 boundary points of $\Omega_3$ falling on the outer sphere $|{\bm n}| = 1$ are 
 $-\frac{1}{\sqrt{3}},\, -\frac{1}{\sqrt{3}},\, \frac{2}{\sqrt{3}}$ 
 (irrespective of whether  
 $\mu = -\frac{1}{\sqrt{3}}$ or $\frac{2}{\sqrt{3}}$) as may be seen from 
(\ref{***2}).  
 But the eigenvalues of ${\bm n}\cdot{\bm \lambda}$ corresponding to 
 boundary points on the inner sphere $|{\bm n}| = 1/2$ (i.e., $\mu = 
(2\sqrt{3})^{-1}$) are 
 $\frac{1}{2\sqrt{3}},\, \frac{1}{2\sqrt{3}},\,- \frac{1}{\sqrt{3}}$. 
Thus, if ${\bm n}\in \mathds{R}^8$ is a boundary point on the 
outer (unit) sphere (i.e., a pure state) then the opposite point of the 
inner sphere, \,$ -\frac{1}{2}{\bm n}$,\, is definitely a boundary point too, and vice 
versa. In other words, {\em boundary points on the inner and outer spheres 
occur in dual pairs}. While the boundary points on the outer sphere 
correspond to vanishing von Neumann entropy, those on the inner sphere correspond to  
maximum entropy (one bit) among all boundary points.

There exists another special sphere  `in between' the inner and outer 
spheres  which happens to be  self-dual.      
 Boundary points on this sphere correspond to those ${\bm n}\in \mathds{R}^8$ 
  for which the matrix ${\bm n}\cdot{\bm \lambda}$ itself is singular  
(in addition to $\mathds{1} +\sqrt{3}\,{\bm n}\cdot{\bm \lambda}$ 
being singular).  We see from 
(\ref{***2}) that this situation corresponds to  
$\mu =  0$ or $\frac{1}{\sqrt{3}}$. 
Interestingly, the `midpoint' of these two values, namely $(2\sqrt{3})^{-1}$,   
is  the value  of $\mu$ that defines the inner sphere. The 
eigenvalues of 
  ${\bm n}\cdot{\bm \lambda}$ for such boundary points are clearly 
 $\frac{1}{\sqrt{3}},\, 0,\, -\frac{1}{\sqrt{3}}$, and so the norm $|{\bm n}| 
 = 1/\sqrt{3}$. It is clear that if ${\bm n}$ corresponds to such a 
boundary point on the sphere of radius $1/\sqrt{3}$ then the antipodal point 
$-{\bm n}$ too is a boundary point. In other words, {\em boundary points on 
the sphere of radius $1/\sqrt{3}$  obtain in pairs of diametrically opposite points}. 
 It is in this sense that this sphere is self dual.   

In our analysis of the two-sections to be taken up in the next Section, 
we shall come across sections of these three 
 special eight-dimensional spheres.  

\section{Two-sections of $\Omega_3$}
That $\Omega_3$ and $\partial \Omega_3$ are substantially richer than
the traditional Bloch ball and its boundary $S^2$ of  a  qubit is transparent. 
 To obtain a feel for this richness it is sufficient to try the challenge of 
visualizing how the four-parameter family $\Omega_3^{\rm ext}$ is 
`sprinkled' over the seven-parameter surface $\partial \Omega_3$ of $\Omega_3$.   
It is in order to gain some insight into the geometry of 
 this `Bloch sphere' $\Omega_3$ 
that we now move on to look at its two and three-dimensional
sections.

 A general two-section of $\Omega_3$ should rightfully mean all density 
operators 
of the
form
\begin{equation}
\rho = \frac{1}{3}\left(\mathds{1} +\alpha \tilde{\lambda}_1 + \beta
  \tilde{\lambda}_2\right), 
\end{equation}
where $ \tilde{\lambda}_1,\,\tilde{\lambda}_2$ is a pair of linearly
independent traceless hermitian matrices, and $\alpha,\,\beta$ are
real. Two-sections determined by
$(\tilde{\lambda}_1,\,\tilde{\lambda}_2)$ and
$(\tilde{\lambda}'_1,\,\tilde{\lambda}'_2)$  are unitarily equivalent
if there exists $U\in SU(3)$ such that the real linear span of
$(U\tilde{\lambda}_1U^{\dagger},\,U\tilde{\lambda}_2U^{\dagger})$ is
the same as that of $(\tilde{\lambda}'_1,\,\tilde{\lambda}'_2)$. 
 It is true that it 
suffices to consider only unitarily inequivalent two-sections, 
 but the manifold  of such two-sections is a huge family: for while we may
choose without loss of generality  $\tilde{\lambda}_1 =
(\cos\theta\lambda_3 + \sin\theta\lambda_8)$, $\tilde{\lambda}_2$  is
then left in the general form
\begin{eqnarray*}
\tilde{\lambda}_2  = (\sin\theta\lambda_3 - \cos\theta\lambda_8) +
\sum_{j\ne 3,8}r_j\lambda_j,
\end{eqnarray*}
where $j$ runs over all the six `off-diagonal' $\lambda$-matrices. The
unitarily inequivalent two-sections are thus parametrized by 
continuous real parameters $\theta,~r_j$ (six in all, since one can be normalized 
away). In order to gain quick partial insight into the geometry of $\Omega_3$,
we consider here not general two-sections, but only standard
two-sections, i.e, sections spanned by a pair of standard
$\lambda$-matrices. 

Such two-sections of 
the quantum state space have been considered earlier by other 
authors\,\cite{kimura,Kossakowski2003,mendas,jakobczyk,kryszewski}.
 Yet we consider them here, briefly, for three reasons: 
(i)~as preparation towards our detailed consideration 
of three-sections in the next Section; (ii)~to point out the special significance 
of the 
circle of radius $1/\sqrt{3}$, and (iii)~to examine the unitary 
equivalence or otherwise of geometrically equivalent sections. 
 
\begin{table}[h]
\begin{center}
\begin{tabular}{|c|c|c|c|}
\hline
Circle&Triangle&Parabola&Ellipse\\
\hline
\{12\}, \{13\}, \{23\}, \{14\}, \{15\}&\{18\}&\{34\}&\{48\}\\
\{16\}, \{17\}, \{24\}, \{25\}&\{28\}&\{35\}&\{58\}\\
\{26\}, \{27\}, \{45\}, \{46\}&\{38\}&\{36\}&\{68\}\\
\{47\}, \{56\}, \{57\}, \{67\}&&\{37\}&\{78\}\\
\hline
\end{tabular}
\end{center}
\caption{
%\footnotesize
The $28$ standard two-sections arranged 
according to their
  types, with $\{jk\}$ denoting the section spanned by the
  $\lambda$-matrices $(\lambda_j , \lambda_k)$.}\label{table-1}
\end{table}
The  $^8C_2 = 28$ two-sections of $\Omega_3$, the state space of 
qutrit, partition  into four
distinct families as shown  in Table~\ref{table-1}.
 To exhibit the shape of each one of the two-sections it is sufficient 
to
obtain an expression for its (closed) boundary, and this is obtained by
restriction of the general expression for 
the closed boundary $\partial
\Omega_3 \subset \mathds{R}^8$, given in Eq.\,(\ref{***}),  
to the two-section under consideration. Restriction to the
$\{12\}$ section gives
\begin{equation}
3(n_1^2+n_2^2) =1,
\end{equation}
which is a circle of radius $1/\sqrt{3}$. 
 Since pure states correspond to
$|{\bm n}| =1$,  there is no pure state on this circle of radius
$1/\sqrt{3}$. Restriction of Eq.\,(\ref{***}) to the $\{18\}$  section 
gives
\begin{equation}
3(n_1^2+n_8^2) - 6n_8\left(n_1^2 - \frac{n_8^2}{3}\right) =1,
\end{equation}
which factors into the transparent form 
\begin{equation}
(1+n_8+\sqrt{3}n_1)(1+n_8-\sqrt{3}n_1)(1-2n_8) =0,
\end{equation}
a triangle whose sides are decided by the vanishing of one of the
three linear factors. 
 The three vertices, the points at which two of
the linear factors vanish simultaneously, are: $(n_1,~n_8)=
(0,\,-1),~\frac{1}{2}(\pm \sqrt{3},\,1)$. Note that
$|{\bm n}|^2 = n_1^2+n_2^2 = 1$ for all  three points, showing that 
these
vertices correspond to pure states. The corresponding three Hilbert space
vectors $|\psi\rangle$ are 
respectively $(0,\,0,\,1),~(1,\,0,\,0),~(0,\,1,\,0)$.
Restriction of Eq.\,(\ref{***}) to the $\{34\}$ section yields 
\begin{equation}
3(n_3^2+n_4^2) -3\sqrt{3}n_3n_4^2 =1,
\end{equation}
which factors into
\begin{equation}
(1+\sqrt{3}n_3 -3n_4^2)(1-\sqrt{3}n_3) =0,
\end{equation}
a parabola $1+\sqrt{3}n_3 =3n_4^2$,  truncated along the line $n_3 =
1/\sqrt{3}$,  this line meeting the parabola at $(n_3,n_4) =
\left(\frac{1}{\sqrt{3}},\pm \sqrt{\frac{2}{3}}\right)$. We find
$|{\bm n}|^2 = n_3^2 + n_4^2 = 1$ at these two points, showing that 
these points
correspond to pure states. 
 The corresponding Hilbert space vectors $|\psi\rangle$ are 
 $\frac{1}{\sqrt{3}} (\sqrt{2},\,0,\,\pm1)$.
 Finally, restriction of the general expression Eq.\,(\ref{***}) for
$\partial \Omega_3$ to the $\{48\}$ section results in 
\begin{equation}
3(n_4^2+n_8^2) -3n_8n_4^2 =1,
\end{equation}
which factors into
\begin{equation}
\left( 3n_4^2 + 2\left(n_8+\frac{1}{4}\right)^2 -
  \frac{9}{8}\right)(1+n_8) =0,
\end{equation}
an ellipse with major axis along the $n_8$ axis, semimajor axis $3/4$,
semiminor axis $\sqrt{3/8}$, and centred at $(x_4,x_8) =
(0,-1/4)$. 
The only pure state on this ellipse is at $(n_4,n_8) =
(0,-1)$, the corresponding Hilbert space vector being
 $|\psi\rangle = (0,\,0,\,1)$
\begin{figure}
\begin{center}
\includegraphics[width=10cm]{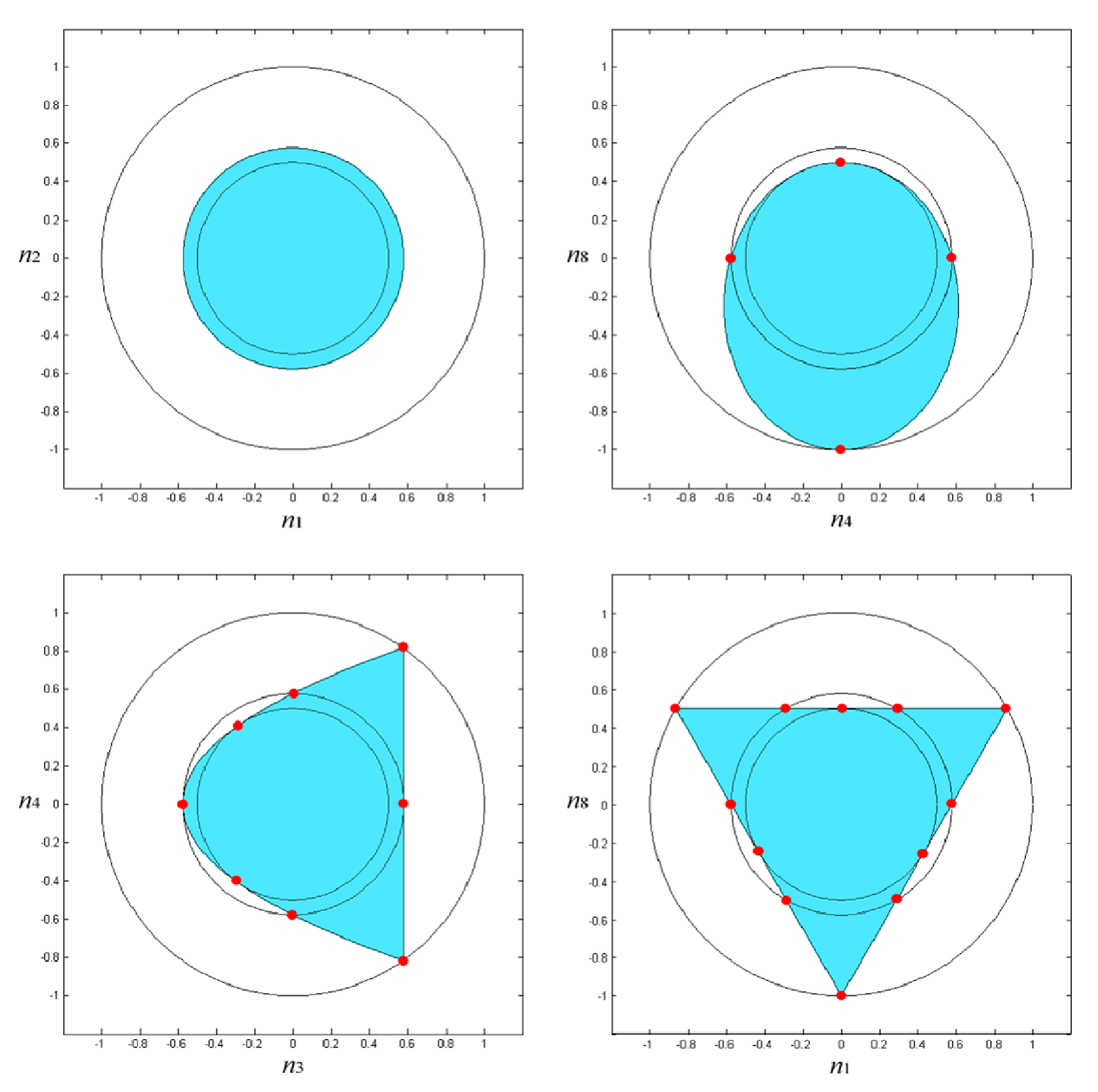}
\end{center}
 \caption{\footnotesize{The four types of two-sections of $\Omega_3$along with  the circles of radii $1$, $1/\sqrt{3}$, and $1/2$ these being sections 
of the three special spheres discussed in Section~3C. The triangular, parabolic, and 
elliptic boundaries are seen to touch the `inner' and `outer' spheres at respectively 
three, two, and one dual pairs of points, and the `middle' sphere at correspondingly 
equal number of self-dual pairs. }}\label{two-secs}
\end{figure}

The two-sections thus assume one of the four closed convex 
shapes---triangle, (truncated) parabola,  
ellipse, or circle---exhibited in Fig.\,\ref{two-secs}.   
 While the first three have respectively $3,\,2,\,1$ pure states, the last 
two-section has none. {\em Since unitary transformations $\rho \to U\rho 
U^{\dagger},\;U\in SU(3)$ are $SO(8)$ rotations on $\Omega_3$, unitarily 
equivalent sections are geometrically equivalent. But the converse is not 
necessarily true (not all $SO(8)$ rotations correspond to $SU(3)$ conjugation), 
and therefore we now turn to consideration of the issue of unitarily equivalent 
sections. }

Unitary equivalence of the three triangular two-sections
$\{18\},\,\{28\},\,\{38\}$ is easily seen: conjugation by the unitary
matrix $\exp\left(i\frac{\pi}{4}\lambda_j\right)$ takes $\lambda_k$ to
$\pm \lambda_l$ (this signature is of no consequence for our purposes)
without affecting $\lambda_8$. Here $jkl = 123$, or a
permuted version thereof. Turning to the four parabolic two-sections, we
see that conjugation by $U={\rm diag}(1,\,1,\,i)$ takes $\lambda_4$ to
$\lambda_5$ and $\lambda_6$ to $\lambda_7$, leaving $\lambda_3$
invariant, thus establishing  the unitary equivalences
$\{34\}\sim\{35\},\,\{36\}\sim\{37\}$.  Conjugation by
$\exp\left(i\frac{\pi}{2}\lambda_2\right)$ takes $\lambda_4$ to
$\lambda_6$ and $\lambda_5$ to $\lambda_7$ leaving $\lambda_3$
essentially unaffected (actually $\lambda_3$ is transformed to
$-\lambda_3$, but for our purpose on hand this is as good as
$\lambda_3$ being left invariant), showing the unitary
equivalences $\{34\}\sim \{36\},~\{35\}\sim\{37\}$,  and thus demonstrating 
that 
the four parabolic sections are indeed unitarily equivalent to one 
another.  
 Unitary equivalence of
the four elliptic two-sections may be seen in {\em exactly} the same 
manner as the parabolic case.

The remaining seventeen two-sections are all circles of the same radius
$1/\sqrt{3}$, as may be seen through restriction of Eq.\,(\ref{***}). 
This
by itself, however, does not prove their unitary equivalence, as 
noted earlier. Indeed,
{\em these circular sections are not in the same unitary equivalence 
class}. And so this case
deserves a more careful examination.

Note, first of all, that the $\{12\},~\{13\}$, and $\{23\}$ sections are
unitarily equivalent to one another, for
$\exp\left(i\frac{\pi}{4}\lambda_j\right)$ leaves, under conjugation,
$\lambda_j$ invariant and rotates $\lambda_k$ and $\lambda_l$. Here
$jkl = 123$ or a permuted version of the same. Further, conjugation by
$\exp\left(i\frac{\pi}{2}\lambda_7\right)$ takes
$\lambda_1,\,\lambda_2$ to $\lambda_4,\,\lambda_5$ and that by
$\exp\left(i\frac{\pi}{2}\lambda_5\right)$ takes
$\lambda_1,\,\lambda_2$ to $\lambda_6,\,\lambda_7$, showing the
equivalences $\{12\}\sim\{45\}\sim\{67\}$.

We have thus shown that the $\{12\},~\{13\},~\{23\},~\{45\}$, and $\{67\}$ sections,
forming a five-element subset, are unitarily equivalent to one
another. Unitary equivalence of the remaining twelve elements
$\{14\}, \{15\}, \{16\}, \{17\}, \{24\}, \{25\}, \{26\}, \{27\}, \{46\}, \{47\}, \{56\}$,
and $\{57\}$ can be similarly seen using conjugation by one of the
diagonal unitaries diag$(1,1,i)$, diag$(1,i,1)$, diag$(i,1,1)$ or one
of $\exp\left(i\frac{\pi}{2}\lambda_2\right)$,
$\exp\left(i\frac{\pi}{2}\lambda_5\right)$,
$\exp\left(i\frac{\pi}{2}\lambda_7\right)$.

But these two subsets of circular two-sections form different unitary 
 equivalence classes. To prove this it
suffices to simply point out that {\em the first subset consists of 
anti-commuting
pairs of $\lambda$-matrices, a property the second subset does not
share}. Equivalently, $ (\,{\bm n}\cdot {\bm \lambda}\,)^2$ are simultaneously 
diagonal for all ${\bm n}$  in the first case. But this is not true of the second 
class, since  $ (\,{\bm n}\cdot {\bm \lambda}\,)^2$  for different ${\bm n}$'s 
do not commute for this class. 

The fact that there are two unitarily inequivalent classes of circular 
two-sections does not seem to have been appreciated in  earlier 
 considerations of  
two-sections\,\cite{kimura,Kossakowski2003,mendas,jakobczyk,kryszewski}.

\vskip 0.3cm
\noindent
{\bf Remark}\,: Care should be exercised in comparing our results on two-sections with 
those of Kryszewski and Zachcial\,\cite{kryszewski}. They have more than four  
rotationally 
inequivalent sections. The reason for this departure is not hard to 
see. They have made an unusual choice for normalization of their $\lambda$-matrices 
in that ${\rm tr} \,\lambda_j^{\,2}$ is not the same for all values of $j$. 
As a consequence, the adjoint action $\rho \to U\rho U^\dagger$ of $U\in SU(3)$ 
{\em does not act as rotation on their $\Omega_3$}.  This in turn forces the 
situation 
wherein unitarily equivalent two-sections are not necessarily rotationally equivalent!  

In Fig.\,\ref{two-secs} depicting the four geometrically different shapes of 
two-sections we have shown also the unit circle and the circles of radii 
$1/\sqrt{3},\,1/2$.  As was to be expected, the boundary of no 
two-section strays into the circle of radius $1/2$ or outside the unit 
circle. Further, if the boundary touches the circle of radius $1/2$ at 
some point, at the diametrically opposite point it touches the 
unit circle. In other words, such boundary points always occur 
in dual pairs: there are three such dual pairs for the triangular 
section, two for the parabolic, one for the elliptic, and none for the 
circular section. 
 
While the unit circle and the circle of radius $1/2$ are mutually dual 
in this sense, the circle of radius $1/\sqrt{3}$ is self-dual: if a 
point of this circle is a boundary point, its antipodal point will also 
be a boundary point. It can be seen that there are three such self-dual antipodal 
pairs for the triangular section, two for the parabolic, and one for the 
elliptic section; the entire circular section comprises, of course, only 
such pairs.
   
\section{Three-sections of $\Omega_3$}
In the last Section we considered briefly the manner in which the $28$
standard two-sections of $\Omega_3$ get themselves organized into four
families or geometric types. We shall now extend this analysis to the 
 richer case of the $^8C_3 =
56$ standard three-sections. As will be seen below, there are seven
 geometrically distinct types of three-sections, and  
Table\,\ref{table-2} 
shows the type to
which each of these $56$ sections belongs. 
\begin{table}
\begin{center}
\begin{tabular}{|c|c|c|c|c|c|c|}
\hline
Sphere&Ellipsoid&Cone&Obese      &{RS1}&{RS2}&Paraboloid\\
      &  &    &Tetrahedron&     &     &\\
\hline
\hline
\{123\}, \{245\}&\{458\}&\{128\}&\{146\}&\{134\}&\{148\}&\{345\}\\
\{124\}, \{246\}&\{468\}&\{138\}&\{157\}&\{135\}&\{158\}&\{367\}\\
\{125\}, \{257\}&\{478\}&\{238\}&\{247\}&\{136\}&\{168\}&\\
\{126\}, \{267\}&\{568\}&\{348\}&\{256\}&\{137\}&\{178\}&\\
\{127\}, \{456\}&\{578\}&\{358\}&\{346\}&\{234\}&\{248\}&\\
\{145\}, \{457\}&\{678\}&\{368\}&\{347\}&\{235\}&\{258\}&\\
\{147\}, \{467\}&&\{378\}&\{356\}&\{236\}&\{268\}&\\
\{156\}, \{567\}&&&\{357\}&\{237\}&\{278\}&\\
\{167\}&&&&&&\\
\hline
\hline
\end{tabular}
\end{center}
 \caption{\footnotesize{The 56 three-sections arranged according to 
their geometric types, $\{jkl\}$ denoting the section spanned by  
$(\lambda_j,\,\lambda_k,\,\lambda_l)$.}}\label{table-2}
\end{table}

In analyzing each type of three-sections we shall
consider one member in some detail, display the density operators 
pertaining to the member, and 
then 
indicate {\em if and how} all the
other members of the same geometric type are unitarily equivalent 
to the section
considered. The three-sections spanned by
$\left(\lambda_j,\,\lambda_k,\,\lambda_l\right)$ will be simply 
denoted
$\{jkl\}$  and, naturally,  our strategy will be to take 
Eq.\,(\ref{***})   
describing
 the closed boundary $\partial \Omega_3$ of our Bloch ball $\Omega_3$ and restrict it 
to the linear span
of  $\left(\lambda_j,\,\lambda_k,\,\lambda_l\right)$ to obtain a
 description of the closed boundary of the section $\{jkl\}$. We begin 
our analysis with the case of cone.
\subsection{Cone}
\begin{figure}
\begin{center}
\includegraphics[width=9.5cm]{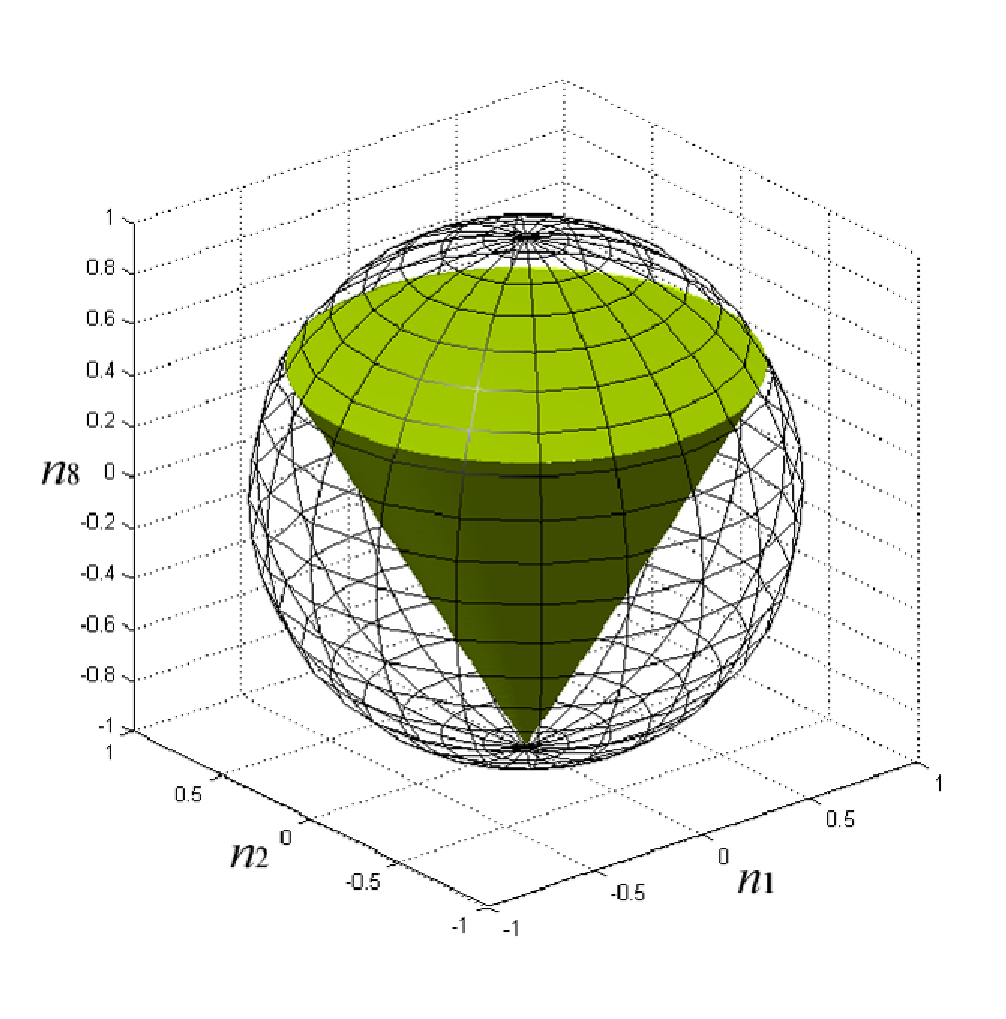}
\end{center}
 \caption{\footnotesize{The conical three-section \{128\}   
 having a singleton pure state at its vertex and a circle of pure 
states at its base.}}\label{cone}
\end{figure}

The particular three-section we shall consider for the conical type is
$\{128\}$, spanned by the triplet
$\left(\lambda_1,\,\lambda_2,\,\lambda_8\right)$. Restriction of
Eq.\,(\ref{***}) to the three space $\{128\}$ reads  
\begin{equation}
3(n_1^2+n_2^2+n_8^2) -6n_8\left(n_1^2+n_2^2 -\frac{n_8^2}{3}\right)=1.
\end{equation}
This equation readily factors into the suggestive form
\begin{equation}
\left(\,(n_8+1)^2 - 3(n_1^2 + n_2^2)\,\right)(2n_8-1) =0,
\end{equation}
which is clearly a cone with vertex at $(n_1,\,n_2,\,n_8) = 
(0,\,0,\,-1)$,
truncated by the base plane $n_8~=~1/2$, as shown in
Fig.\,\ref{cone}. There is a singleton pure state at $(0,\,0,\,-1)$
and a circle of pure states at
$n_8~=~1/2,~\sqrt{n_1^2+n_2^2}~=~\sqrt{3}/2$; these are the points at
which the cone touches the outer sphere (unit sphere). 

Clearly, the density operators belonging to this three-section are
of the form
\begin{equation}
\rho =\frac{1}{3}\left(
\begin{array}{ccc}
1+n_8&\sqrt{3}(n_1-in_2)&0\\
\sqrt{3}(n_1+in_2)&1-n_8&0\\
0&0& 1-2n_8
\end{array}
\right),
\end{equation}
 which can be verified to be positive if and only if 
the three-vector ${\bm n}=(n_1,\,n_2,\,n_8)$ is in this
 (truncated) solid cone (here, as in the rest of this Section, 
three-component vectors like ${\bm n} =(n_1,\,n_4,\,n_6)$ 
will always refer to the corresponding 
${\bm n} =(n_1,\,0,\,0,\,n_4,\,0,\,n_6,\,0,\,0)\in \mathds{R}^8$). 

And it can be seen that the 
 singleton and circle of pure states correspond respectively to 
Hilbert space unit vectors $|\psi\rangle =
(0,\,0,\,1)$ and $\frac{1}{\sqrt{2}}(1,\,e^{i\theta},\,0)$.

That the seven three-sections listed under `cone' in 
Table\,\ref{table-2} are
unitarily equivalent to one another is particularly easy to see. 
Conjugation by
$\exp\left(i\frac{\pi}{4}\lambda_1\right)$ leaves
$\lambda_1,\,\lambda_8 $ invariant but transforms $\lambda_2$ to
$\lambda_3$, demonstrating the unitary equivalence $\{128\} \sim
\{138\}$. Similarly, the equivalence $\{138\} \sim \{238\}$ is
established by $\exp\left(i\frac{\pi}{4}\lambda_3\right)$. The 
equivalence $\{138\} \sim \{348\}$ is seen by noting that
conjugation by the unitary permutation matrix
\begin{equation}
U =\left(\begin{array}{ccc}
0&1&0\\
0&0&1\\
1&0&0\end{array}\right)
\end{equation}
takes $\lambda_1$ to $\lambda_4$ and the linear span of
$(\lambda_3,\,\lambda_8)$ onto itself. Finally, the equivalences
$\{348\}\sim \{358\}$ and $\{368\}\sim \{378\}$ are
established by the diagonal unitary matrix ${\rm diag}(1,\,1,\,i)$, 
while $\{348\}\sim \{368\},~\{358\}\sim \{378\}$ 
are witnessed by
$\exp\left(i\frac{\pi}{2}\lambda_2\right)$.
\subsection{Paraboloid}
\begin{figure}
  \begin{center}
 \includegraphics[width=9.5cm]{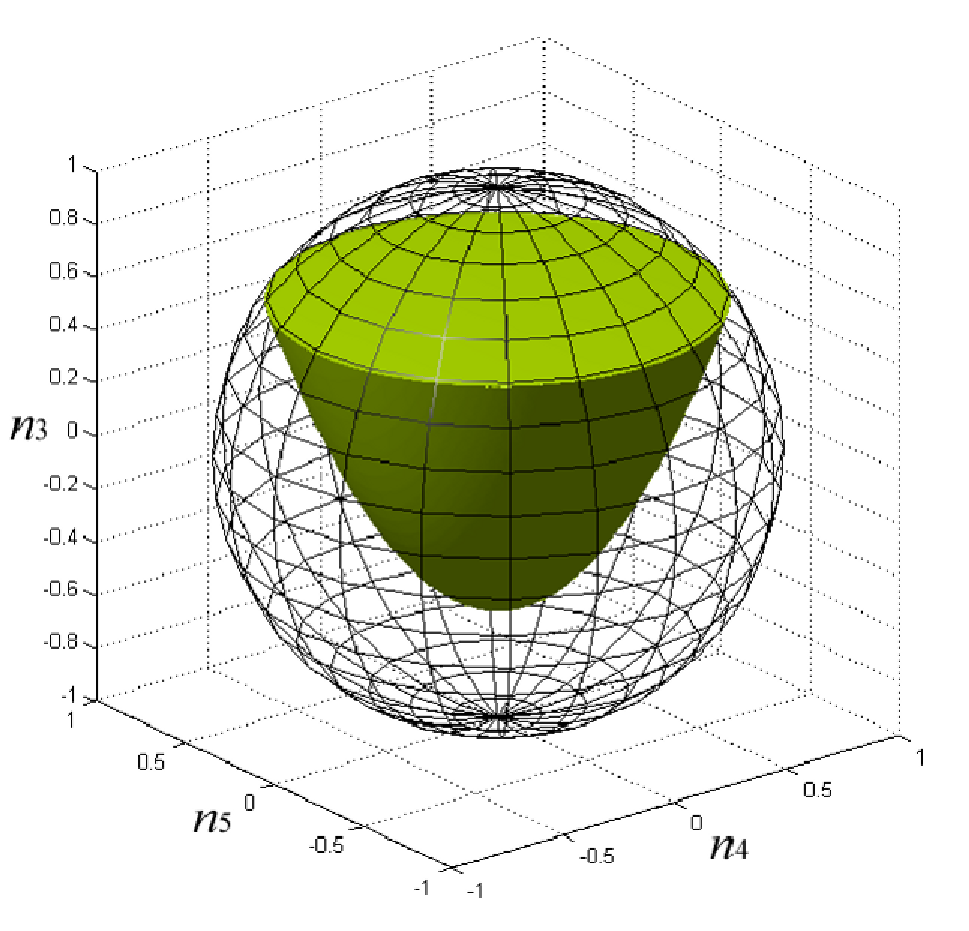}
      \end{center}
\caption{\footnotesize{The paraboloidal three-section \{345\}  
 having a circle of pure states.}}\label{paraboloid}
\end{figure}
The three-sections $\{345\}$ and $\{367\}$ alone  belong to the paraboloid  type. 
Restriction of Eq.\,(\ref{***}) 
 to the $\{345\}$ case reads 
\begin{equation}
3(n_3^2+n_4^2+n_5^2) -3\sqrt{3}n_3\left(n_4^2+n_5^2\right)=1,
\end{equation}
which conveniently factors into
\begin{equation}
(1-\sqrt{3}n_3)\left(3(n_4^2 + n_5^2) - (\sqrt{3}n_3 +1) \right)=0. 
\end{equation}
 The closed  boundary of the three-section $\{345\}$ is thus seen to be the 
paraboloid  
$ 3(n_4^2 + n_5^2) = \sqrt{3}n_3 +1$ truncated by the plane $n_3 =
1/\sqrt{3}$, the latter acting as the `closing lid', as shown in
Fig.\,\ref{paraboloid}. This three-section has a circle worth of pure 
states
 and these correspond to $n_3 = 1/\sqrt{3},~n_4^2 + n_5^2 = 2/3$, the 
points at which the paraboloid touches the outer sphere. 
The density operators corresponding to this section necessarily have
the form
\begin{eqnarray}
\rho &= \frac{1}{3}\left(\begin{array}{ccc}
1+\sqrt{3}n_3 &0 & \sqrt{3}(n_4-in_5)\\
0 & 1-\sqrt{3}n_3 & 0\\
\sqrt{3}(n_4+in_5) &0 &1
\end{array}\right)\!,\;\;\;
\end{eqnarray}
 and we see that positivity of $\rho$ is indeed equivalent to 
the three-vector ${\bm n} = (n_3,\,n_4,\,n_5)$ 
being in the
solid (truncated) paraboloidal region of Fig.\,\ref{paraboloid}. The 
 circle worth of pure
states correspond to  Hilbert space vectors $|\psi\rangle =
\frac{1}{\sqrt{3}}(\sqrt{2},\,0,\,e^{i\theta}),\,~0\le \theta <
2\pi$. Finally the fact that the three-sections $\{345\}$ and $\{367\}$
are unitarily equivalent may be seen through conjugation by
$\exp\left(i\frac{\pi}{2}\lambda_2\right)$.
\subsection{Ellipsoid}
\begin{figure}
  \begin{center}
\includegraphics[width=9.5cm]{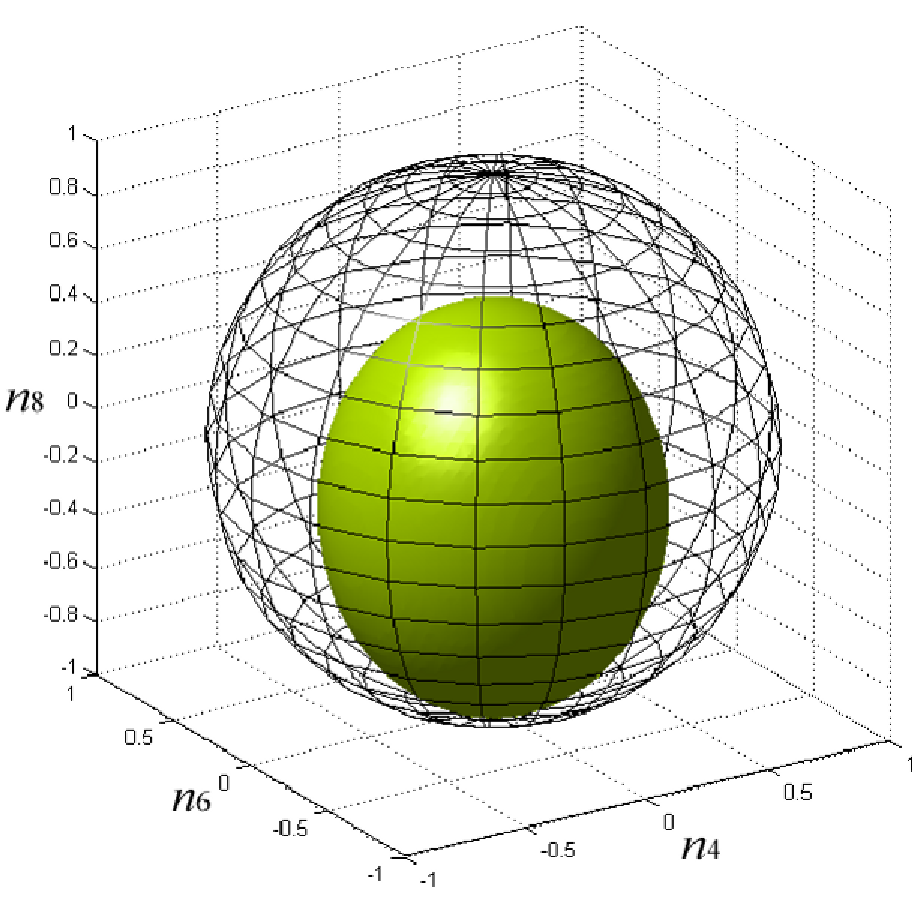}
      \end{center}
 \caption{\footnotesize{The ellipsoidal three-section \{468\} 
having only one pure state.}}\label{ellipsoid}
\end{figure}
As may be seen from Table\,\ref{table-2}, there are six 
three-sections of this
type, and we begin with the specific case of $\{468\}$. 
 Restriction of Eq.\,(\ref{***}) to this case reads
\begin{equation}
3(n_4^2+n_6^2+n_8^2) +n_8\left(3n_4^2+3n_6^2 + 2n_8^2\right)=1.
\end{equation}
This cubic expression factorizes, as in the previous two cases, into
\begin{equation}
 (1+n_8)\left(3(n_4^2 + n_6^2) +2\left(n_8 +\frac{1}{4}\right)^2 
-\frac{9}{8} \right)=0.
\end{equation}
This is clearly an ellipsoid of revolution (about the major axis
$n_8$), with semimajor axis $3/4$ and semiminor axis
$\sqrt{3/8}$, with the centre of the ellipsoid located at
$(n_4,\,n_6,\,n_8) = (0,\,0,\,-1/4)$. It touches the outer sphere only at the 
point $(0,\,0,\,-1)$, as seen also from Fig.\,\ref{ellipsoid} and hence there
is only one pure state in this three-section. As expected, the boundary touches 
the inner sphere at the dual point $(0,\,0,\,1/2)$ and only at this point.  
Any density operator that belongs to the $\{468\}$ section has to
have the form
\begin{equation}
\rho = \frac{1}{3}\left(\begin{array}{ccc}
1+n_8 & 0 & \sqrt{3}n_4\\
0&1+n_8&\sqrt{3}n_6\\
\sqrt{3}n_4&\sqrt{3}n_6 & 1-2n_8
\end{array}\right),
\end{equation}
and we see that $\rho\ge 0$ if and only if 
the three-vector ${\bm n} =(n_4,\,n_6,\,n_8)$ lies in this solid
ellipsoid. The singleton pure state is seen to correspond to  Hilbert
space vector $|\psi\rangle = (0,\,0,\,1)$.

The four components $n_4,~n_5,~n_6,~n_7$ of ${\bm n}$ enters the relevant
part of Eq.\,(\ref{***}) {\em symmetrically}, and only in the 
combination
$n_4^2+n_5^2+n_6^2+n_7^2$. It follows that all the six three-sections
listed under `ellipsoid' in Table\,\ref{table-2} will have identical 
geometric structure, for they are obtained by simply picking two out 
of these four components to partner with $n_8$.  
However, while unitary equivalence of two three-sections
implies their geometric equivalence, the converse is not true in
general, and so we explore this issue. 

Unitary equivalence of $\{468\}$ and $\{478\}$ is established through the
diagonal unitary ${\rm diag}\,(1,\,i,\,1)$. The same is true of the 
equivalence $\{578\} \sim
 \{568\}$ as well. And the unitary equivalence $\{468\} \sim \{578\}$ 
 is 
 established through the unitary ${\rm diag}(1,\,1,\,i)$. Thus, the 
 sections $\{468\},\,\{478\},\,\{578\}$, and $\{568\}$ are unitarily 
equivalent 
to one another. Similarly $\{458\},~\{678\}$ is a
pair of unitarily equivalent sections, as may be seen through
conjugation by $\exp\left(i\frac{\pi}{2}\lambda_2\right)$. {\em These two
 subsets are, however, unitarily inequivalent}. Proof consists in 
 simply noting that $\lambda_4,\,\lambda_5$ (as also 
$\lambda_6,\,\lambda_7$) anticommute, but there 
 exists no such pair in the (real) linear span of 
$\lambda_4,\,\lambda_6,\, \lambda_8$.

\subsection{Obese-tetrahedron}
The conical, paraboloidal, and ellipsoidal three-sections considered 
so far are related respectively to the triangular, parabolic, and elliptical 
two-sections of the previous Section in an obvious manner. It is for this reason 
that the cubic expression forming the left hand side of Eq.\,(\ref{***}) factorized   
into a quadratic and a
linear expression in these three cases. It turns out that the next three 
 types of three-sections are {\em genuinely cubic}. 

\begin{figure}
\begin{center}
\includegraphics[width=9.5cm]{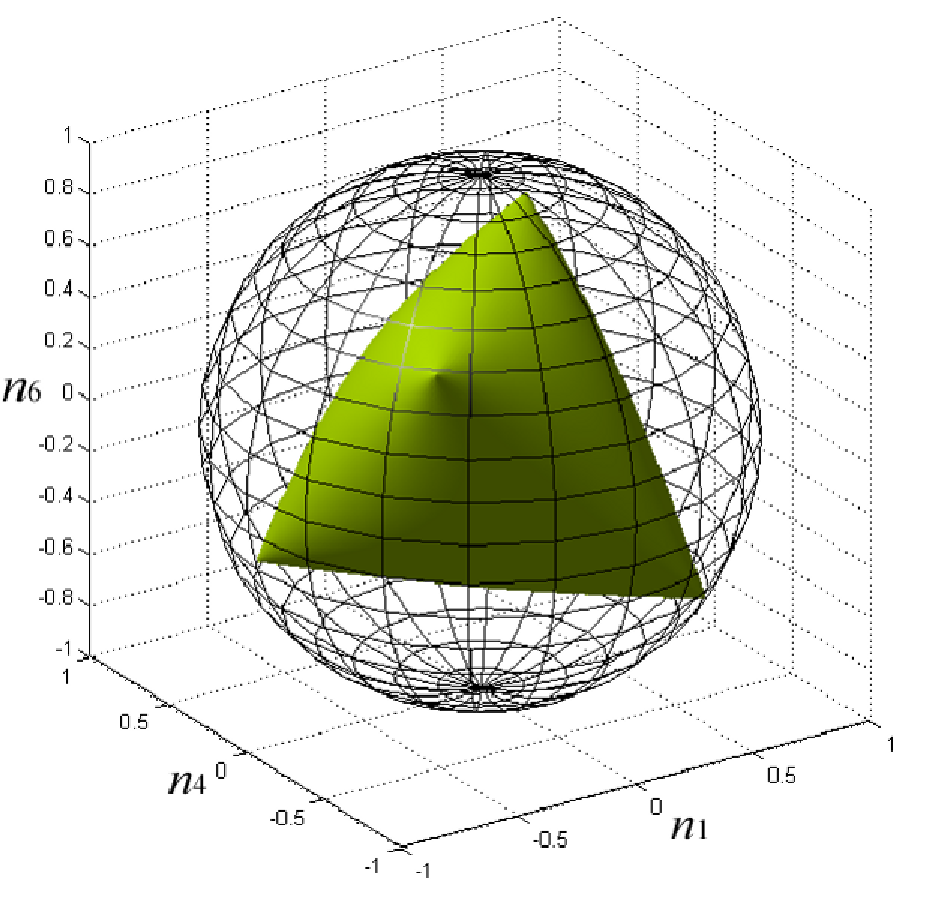}
 \end{center}
 \caption{\footnotesize{The 
obese-tetrahedral three-section~\{146\} 
 having four pure states.}}\label{obese-tetrahedron}
\end{figure}
There are eight three-sections of 
the obese-tetrahedral type as shown in Table\,2, and we begin 
with the
case of $\{146\}$. Restriction of Eq.\,(\ref{***}) 
 which describes  the boundary of
$\Omega_3$  to the present case gives the following equation for the 
boundary of the $\{146\}$ section:
\begin{equation}
3(n_1^2+n_4^2+n_6^2) -6\sqrt{3}n_1n_4n_6=1.\label{YYY}
\end{equation}
{\em Unlike the previous three cases, this expression does not
 factorize:  the boundary of this three-section is genuinely cubic}.  
 For reasons which will become clear later on, we call this three 
 dimensional solid, shown in Fig.\,\ref{obese-tetrahedron}, the {\em 
obese-tetrahedron}. 

 This section has four, and only four, pure states, the points at 
which the obese-tetrahedron touches the outer sphere. These are 
\begin{eqnarray}
\hspace{-2.5cm}(n_1,\,n_4,\,n_6) =&&
\frac{1}{\sqrt{3}}(1,\,1,\,1),~\frac{1}{\sqrt{3}}(1,\,-1,\,-1),~\frac{1}{\sqrt{3}}(-1,\,1,\,-1),~\frac{1}{\sqrt{3}}(-1,\,-1,\,1).\;\;\;
\end{eqnarray}
 The density operators corresponding to this section necessarily 
assume 
the form
\begin{equation}
\rho = \frac{1}{3}\left(\begin{array}{ccc}
1&\sqrt{3}n_1&\sqrt{3}n_4\\
\sqrt{3}n_1&1&\sqrt{3}n_6\\
\sqrt{3}n_4&\sqrt{3}n_6&1
\end{array}\right),
\end{equation}
 which is verified to be positive if and only if 
${\bm n}=(n_1,\,n_4,\,n_6)$ is a point in
the obese
tetrahedron shown in Fig.\,\ref{obese-tetrahedron}. It is
readily seen that  Hilbert space vectors $|\psi\rangle$ corresponding
to the above four pure states are, respectively, $
\frac{1}{\sqrt{3}} (1,\,1,\,1),\,\,
\frac{1}{\sqrt{3}} (1,\,1,\,-1),\,\,
\frac{1}{\sqrt{3}} (1,\,-1,\,1),\,\,
\frac{1}{\sqrt{3}} (-1,\,1,\,1)$. 

Unitary equivalence of the eight obese-tetrahedral sections can be 
seen as
follows. The unitary equivalences $\{146\} \sim \{157\},\,\{146\}\sim 
\{247\}$,
and $\{146\}\sim \{256\}$ are seen through conjugation by 
diag$(1,\,1,\,i)$,
diag$(1,\,i,\,1)$, and diag$(i,\,1,\,1)$ respectively, showing that 
the
four sections $\{146\},\,\{157\},\,\{247\},\,\{256\}$ are unitarily equivalent 
to
one another. Similarly, the unitary equivalences $\{346\}\sim 
\{347\},\,\{346\}\sim \{356\}$, and $\{346\}\sim\{357\}$ are seen through
conjugation by diag$(1,\,i,\,1)$, diag$(i,\,1,\,1)$ and 
diag$(1,\,1,\,i)$
respectively, showing that the four sections $\{346\},\,\{347\},\,
\{356\},\,\{357\}$ are unitarily equivalent  to one another. Finally, the 
unitary
equivalence $\{146\}\sim \{346\}$ is seen through conjugation by
$\exp\left(i\frac{\pi}{4}\lambda_2\right)$ which takes $\lambda_1$ to
$\lambda_3$, leaving invariant the linear span of
$(\lambda_4,\,\lambda_6)$. This completes proof of  unitary 
equivalence of the eight obese-tetrahedral three-sections to one another.

The next two types to be considered also turn out to be genuine 
cubic
sections. Since these two shapes are unfamiliar, at least to the 
present
authors, these two three-sections will be simply denoted {RS1} and {RS2}.

\subsection{{RS1}}
\begin{figure}
\begin{center}
\includegraphics[width=8.5cm]{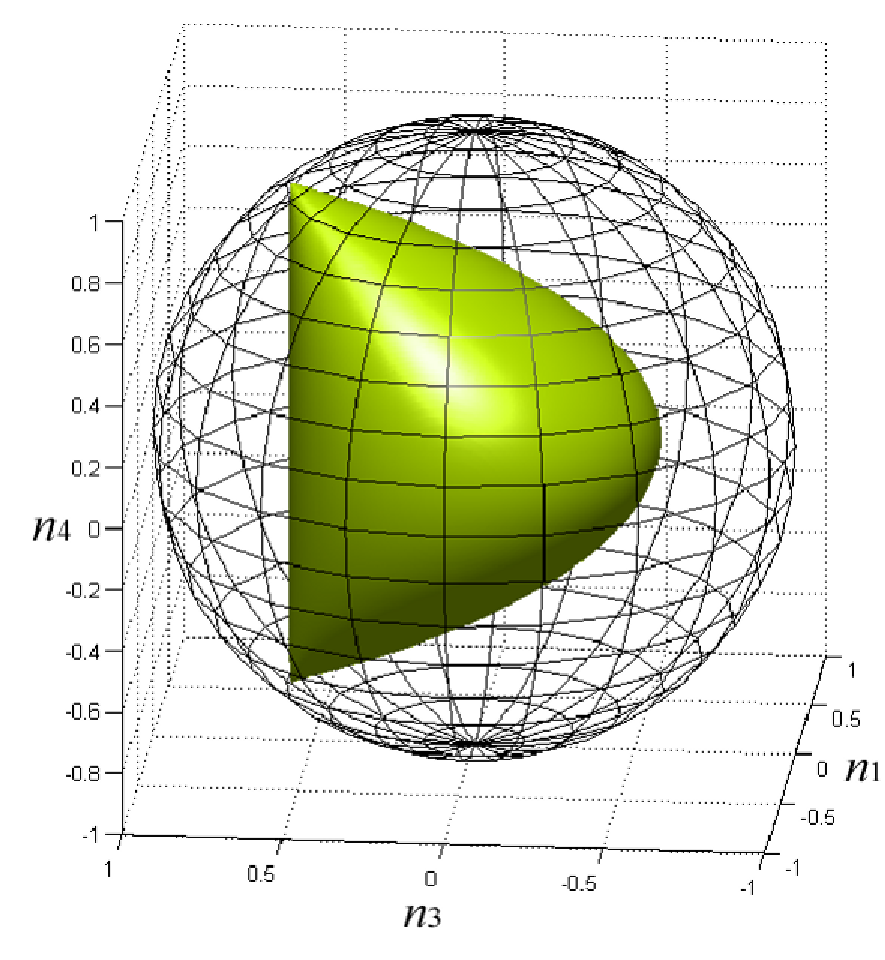}
\end{center}
 \caption{\footnotesize{The {RS1} three-section \{134\} having two pure states.}}\label{RS11}
\end{figure}

As may be seen from Table\,\ref{table-2} there are eight 
three-sections of type
 {RS1}.  We consider first the section $\{134\}$. Restriction of
Eq.\,(\ref{***}) to $\{134\}$ reads
\begin{equation}
3(n_1^2+n_3^2+n_4^2) -3\sqrt{3}n_4^2n_3=1,
\end{equation}
giving the boundary of {RS1}  shown in Fig.\,\ref{RS11}. It is to be
appreciated that the signature in front of $3\sqrt{3}n_4^2n_3$ has no
effect on the shape of the section, for it can be  simply
 absorbed into $n_3$.
This section has
just two pure states, represented by the pair of points
$(n_1,\,n_3,\,n_4) =
\left(0,\,\frac{1}{\sqrt{3}},\,\pm\frac{2}{\sqrt{3}}\right)$ at which
 {RS1} touches the unit  sphere. 
 It is clear that these two pure states
correspond to Hilbert space vectors $|\psi\rangle =
(\sqrt{2},\,0,\,\pm 1)/\sqrt{3}$. 

The density operators of this section 
have to be
of the form
\begin{equation}
\rho =\frac{1}{3}\left(\begin{array}{ccc}
1+\sqrt{3}n_3 & \sqrt{3}n_1 & \sqrt{3}n_4\\
\sqrt{3}n_1 & 1-\sqrt{3}n_3 & 0\\
\sqrt{3}n_4 & 0 & 1
\end{array}\right),
\end{equation}
which is easily verified to be positive if and only if the condition $
3(n_1^2+n_3^2+n_4^2) 
-3\sqrt{3}n_4^2n_3\le 1$ is met, i.e., if and only if 
${\bm n} = (n_1,\,n_3,\,n_4) \in\,${RS1}. 

Unitary equivalence of the eight sections of type {RS1}  can be seen as
 follows. That $\{134\}\sim \{135\}$ follows from conjugation by 
diag$(1,\,1,\,i)$, and $\{136\}\sim \{137\}$  follows from the same
conjugation. The equivalences $\{134\}\sim \{136\},\,\{135\}\sim \{137\}$
follow from conjugation by $\exp\left(i\frac{\pi}{2}\lambda_2\right)$,
thus proving the equivalence of $\{134\},\,\{135\},\,\{136\},\,\{137\}$ to one
another.

The equivalence $\{234\}\sim\{235\}$ follows from conjugation by
diag$(1,\,1,\,i)$, and the same conjugation establishes also the
equivalence $\{236\}\sim \{237\}$. The equivalences $\{234\}\sim \{236\},~
\{235\}\sim \{237\}$ are seen through conjugation by
$\exp\left(i\frac{\pi}{2}\lambda_2\right)$, thus proving equivalence
of $\{234\},\,\{235\},\,\{236\},\,\{237\}$ to one another.

Finally, the equivalences $\{134\}\sim \{234\}$ and $\{135\}\sim \{235\}$ are
seen through conjugation by diag$(1,\,i,\,1)$, and the equivalences
$\{136\}\sim \{236\}$ and $\{137\}\sim \{237\}$ through 
conjugation by diag$(i,~1,~1)$, thus establishing unitary equivalence 
of the eight three-sections of type {RS1}. 

\subsection{{RS2}}
\begin{figure}
\begin{center}
\includegraphics[width=7.5cm]{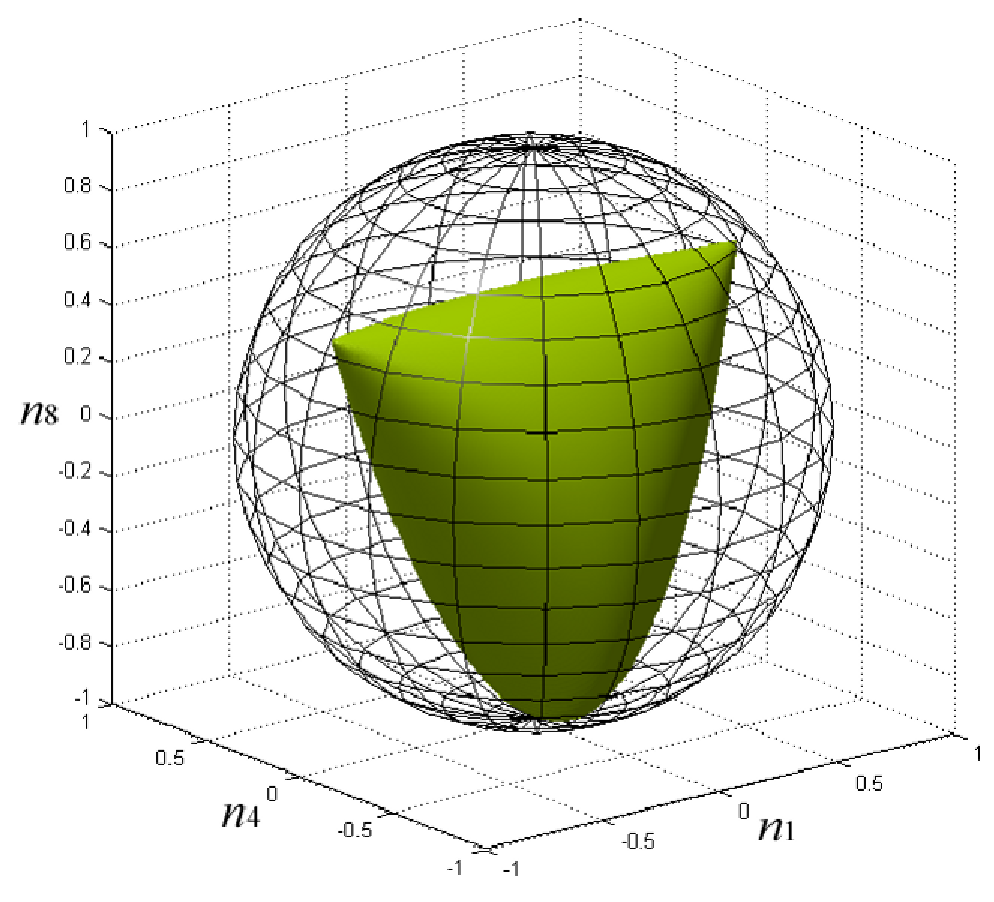}
\end{center}
 \caption{\footnotesize{The three-section  \{148\} 
 of type  {RS2}, having three pure states.}}\label{RS21}
\end{figure}

 There are eight three-sections of this type, and these are listed in 
Table\,\ref{table-2}. We begin with the section $\{148\}$. Restriction 
of
Eq.\,(\ref{***}) to this section reads
\begin{equation}
3(n_1^2+n_4^2+n_8^2) -3n_8\left(2n_1^2 -n_4^2 -\frac{2}{3}n_8^2\right)=1,
\end{equation}
giving the boundary of {RS2}  shown in Fig.\,\ref{RS21}. This section 
has
three pure states corresponding to the points $(n_1,\,n_4,\,n_8) =
 (0,\,0,\,-1),~\left(\pm\frac{\sqrt{3}}{2},\,0,\,\frac{1}{2}\right)$ 
at which {RS2}  touches the unit sphere. 
 Hilbert space vectors $|\psi\rangle$ 
corresponding 
to these three pure states are readily seen to be 
$(0,\,0,\,1),\,\,
  \frac{1}{\sqrt{2}}(1,\,\pm 1,\,0)$.

The density operators which correspond to this section necessarily
assume the form
\begin{equation}
\rho = \frac{1}{3}\left(\begin{array}{ccc}
1+n_8 & \sqrt{3}n_1 & \sqrt{3}n_4 \\
\sqrt{3}n_1 & 1+ n_8 &0 \\
\sqrt{3}n_4 & 0 & 1-2n_8
\end{array}\right),
\end{equation}
whose positivity indeed demands $3(n_1^2+n_4^2+n_8^2) -3n_8\left(2n_1^2
  -n_4^2 -\frac{2}{3}n_8^2\right)\le 1$, i.e, that ${\bm n} = 
(n_1,\,n_4,\,n_8)$ be in {RS2}.

Proof of unitary equivalence of the eight sections of this type follows in
exactly the same way it did in the case of {RS1}. It suffices to note
that (1) these eight sections of type {RS2} are obtained from the
respective eight sections of type {RS1} by simply replacing 
$\lambda_3$ with 
$\lambda_8$, and (2) that every unitary used for conjugation 
 towards proof of unitary equivalence in the  case
of {RS1} mapped $\lambda_3$ to itself, and  it has no effect
on $\lambda_8$.  
\subsection{Sphere}
\begin{figure}
\begin{center}
\includegraphics[width=8.5cm]{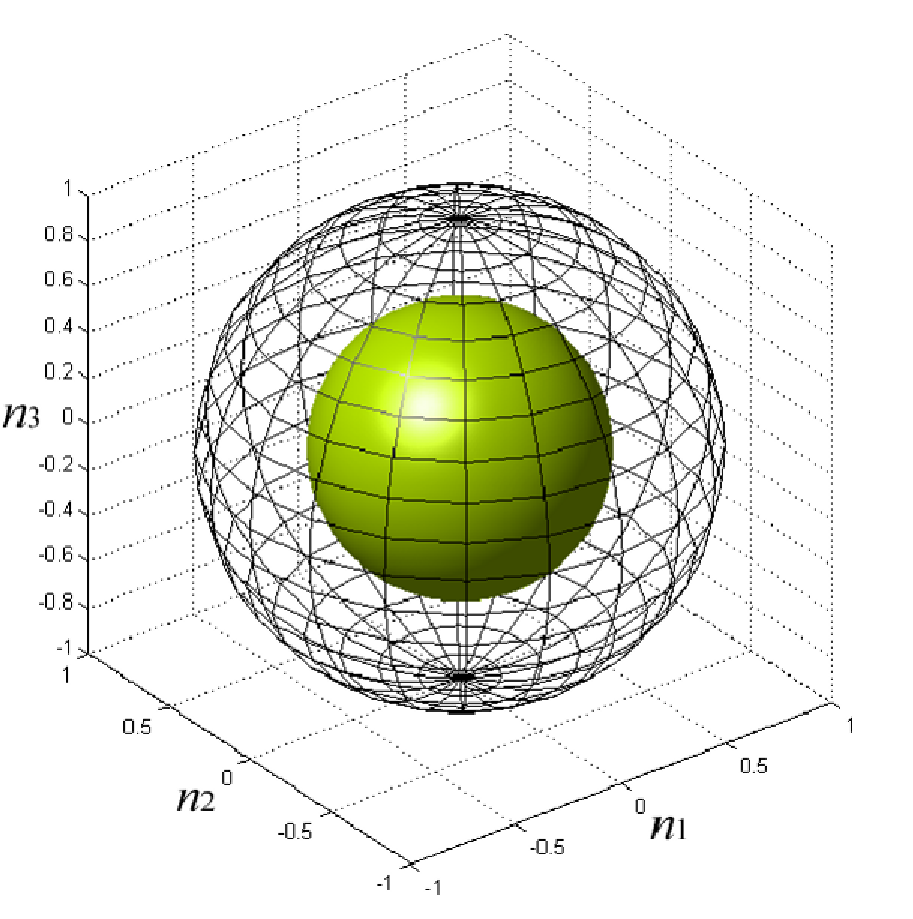}
\end{center}
 \caption{\footnotesize{The three-section \{123\} which is  
spherical, having no pure state.}}\label{sphere}
\end{figure}

As many as seventeen three-sections of $\Omega_3$ are of this
type. These are precisely the three-sections for which the restriction
of Eq.\,(\ref{***}) becomes purely quadratic. Thus, none of these 
sections
involve $\lambda_8$. They tend to avoid $\lambda_3$ as well, with the
exception of the combination $\{123\}$ which does not enter the cubic  
part of Eq.\,(\ref{***}). Further, $\{146\},~\{157\},~\{256\}$, and $\{147\}$
which involve neither $\lambda_3$ nor $\lambda_8$ do not belong to
this type, but to the obese-tetrahedral case. Thus of the $^6C_3 = 20$
three-sections involving neither $\lambda_3$ nor $\lambda_8$ we remove
the above four obese-tetrahedral sections, and add the exceptional 
case of $\{123\}$ to arrive at the count
seventeen.

Since the cubic terms of Eq.\,(15) contribute to none of these 
sections,
restriction of 
Eq.\,(\ref{***}) to section $\{jkl\}$ of this type reads
\begin{equation}
3(n_j^2+n_k^2+n_l^2) = 1,
\end{equation}
a sphere of radius $1/\sqrt{3}$, as  shown in Fig.\,\ref{sphere}. It 
is
the three-section of minimum $3$-volume, not only among the standard
three-sections, but among all three-sections. It is the only type
which has no pure states.

These seventeen geometrically equivalent three-sections are {\em not 
unitarily equivalent to one
another}. They can be grouped into three subsets. The section $\{123\}$
is a singleton set in itself, for we know that
$\lambda_1,\,\lambda_2,\,\lambda_3$ have  the algebraic 
(commutation, anticommutation) properties 
which are similar to those of the Pauli matrices, 
  and there is no other triplet among the
$\lambda$-matrices which has this characteristic property of $SU(2)$ generators.

Similarly we know that the three `imaginary' $\lambda$-matrices
$\lambda_2,\,\lambda_5,\,\lambda_7$ are $SO(3)$ generators. The 
triplets
$(\lambda_1,\,\lambda_4,\,\lambda_7)$, 
$(\lambda_1,\,\lambda_5,\,\lambda_6)$,
and $(\lambda_2,\,\lambda_4,\,\lambda_6)$ are related to the triplet
$(\lambda_2,\,\lambda_5,\,\lambda_7)$ through conjugations by diagonal
unitaries. And there is no other triplet among the $\lambda$-matrices
which possess this structure of $SO(3)$ generators. So these four
triplets form a second subset of unitarily equivalent three-sections. 

 Finally, the three {\em four-sections} $\{1245\}, 
\,\{1267\}$, and $\{4567\}$ form
spheres of radius $1/\sqrt{3}$, as may be seen from restriction of
Eq.\,(\ref{***}); {\em there exists no other four-section with this
  property}.  These three {\em four-sections} are unitarily equivalent 
to one another: the equivalence $\{1245\}\sim \{1267\}$ is seen through 
conjugation by $\exp{(i\frac{\pi}{2}\lambda_2)}$ 
 and  the equivalence $\{1267\}\sim \{4567\}$  
 by $\exp{(i\frac{\pi}{2}\lambda_7)}$.  
Each one of these spherical four-sections leads to four
spherical three-sections, adding to twelve. That these twelve
spherical three-sections forming the third and last subset are
unitarily equivalent to one another is thus established.
 {\em There are thus three unitarily inequivalent sets of spherical 
three-sections}.  

\subsection{Comparison with the work of Mendas}
The $56$ three-sections of $\Omega_3$, the state space of the qutrit,
were earlier studied by Mendas\, \cite{mendas} using {\em Monte Carlo
sampling} method, leading to the conclusion that there are {\em ten} distinct
types of three-sections. In contrast our approach is {\em analytic}
 and hence conclusive, and we
have shown that there are only {\em seven} types. Further, we have presented
for each type analytic expression for the three-section, and this was
not possible in the Monte Carlo approach. Finally, we have shown, both in
the ellipsoidal and spherical cases, that there are three-sections
which are geometrically the same but nevertheless are unitarily
inequivalent, an aspect which is clearly beyond the purview of the
 Monte Carlo approach.  
 The findings of Mendas\,\cite{mendas} group all the 
six ellipsoidal sections into one `type' (Type~10 of Mendas) and all the 
seventeen spherical three-sections into one type (Type~1); it groups the eight 
conical sections into three types (Types~2,\,8,\,9) and the eight obese-tetrahedral 
sections into two types (Types~4,\,7), but now we know that all the eight conical 
sections are mutually equivalent not only geometrically but also unitarily. We have 
shown also that the same is true of the eight obese-tetrahedral sections as well.  
  
\section{Obese-tetrahedron: A deeper look}
In this Section we examine in more detail the interesting case of  obese-tetrahedral
three-sections. To be specific we consider again the $\{146\}$
section. Several questions may arise from the very name, `obese' and
`tetrahedron': (1) Do these three-sections really possess the full tetrahedral
symmetry $T_d$ familiar from the context of point groups? (2) If the
answer is in the affirmative, where does this symmetry originate from?
And, finally, (3) what forces the tetrahedron to be obese?

We begin with the last question. That {\em a tetrahedral section  of
$\Omega_3$ has to necessarily be obese} can be traced to the very 
geometry
of $\Omega_3$. We have already noted that a boundary point of
$\Omega_3$ which is closest to the origin can be no nearer than a
 distance of $1/2$, the radius of the inner-ball, and that such a 
nearest point will always be
 directly opposite to a pure state (and we know that the latter is at
unit distance from the origin). Now in a tetrahedron with vertices on
the outer (unit) sphere, points closest 
to the origin are at a distance $1/3$; there are four such points, the
 centres of the four faces of the tetrahedron. Thus {\em our 
tetrahedron is obese by the minimal
 amount so that these four `base points' of the tetrahedron at a distance $1/3$ 
from the 
origin 
move away radially to 
$1/2$,   just sufficient to fit the geometry of $\Omega_3$.}

The first two questions turn out to be connected, and so they are best
treated together. Recall that the obese-tetrahedral $\{146\}$ section 
consists of all $3\times 3$ density matrices which are {\em real} with
{\em equal entries along the diagonal}. We ask: what are the $SU(3)$
matrices $U$ which under conjugation $\rho({\bm n})\to\rho'({\bm n})
= \rho({\bm n}') = U\rho({\bm n}) U^{\dagger}$
 map the $\{146\}$ section onto itself. Such matrices will of course
constitute a subgroup of $SU(3)$ and, indeed, it will be a subgroup of
$SO(3) \subset SU(3)$. Since the relevant $SU(3)$ matrices are real,
they will map the $\{257\}$ three-section, the real linear span of the
`imaginary' 
$\lambda$-matrices  onto itself. It follows that the $\{38\}$
two-section of diagonal $\lambda$-matrices too will be mapped onto 
itself under
the conjugation action of these matrices $\in SO(3)\subset SU(3)$. 
 {\em This implies that the relevant $SO(3)$ matrices have to be 
signed or generalized permutation matrices}.  

It is well known
that every $SU(3)$ matrix acts on the eight $\lambda$-matrices, through
conjugation, as an $SO(8)$ rotation; this is the adjoint
representation of $SU(3)$. Thus, our consideration in 
the above  paragraph shows
that the question of what is the subgroup of $SU(3)$ which maps the
$\{146\}$ section in $\mathds{R}^8$ onto itself is the same as the
following question: What is the (maximal) subgroup of $SU(3)$ matrices
whose adjoint representation has the form of a direct sum $3\oplus
3\oplus 2$, the two three-spaces arising as the linear spans
$\{146\}$, $\{256\}$ respectively, and the two-space arising as the linear span
$\{38\}$? This subgroup of real $SU(3)$ matrices is easily determined
and it has $24$ elements: the six $3\times 3$ permutation matrices 
have {\em three} nonzero entries each; choice of independent signature for 
each entry results in $6\times2^3=48$ matrices, exactly half of them have 
positive determinant and hence are elements of $SO(3)$. This 24-element discrete 
subgroup of $SO(3)$  turns out to be 
 {\em isomorphic to the tetrahedral
  group $T_d$} whose character table is reproduced in 
Table\,\ref{table-3}. The pair of three-dimensional irreducible representations marked $T_1,~ T_2$
and the two-dimensional one marked $E$ will prove significant for our purpose.

\begin{table}[h]
\begin{center}
\begin{tabular}{c|crrrr}
$\,T_d\,$&$\,\,[\,e\,]\,$&$\,8\,[\,C_3\,]\,$&$\,3\,[\,C_2\,]\,$
      &$\,6\,[\,S_4\,]\,$&$\,6\,[\,\sigma_d\,]\,$\\
\hline
\hline
$A_1\,$&$1\,$&$1~$&$1~$&$1~$&$1~$\\
$A_2\,$&$1\,$&$1~$&$1~$&$-1~$&$-1~$\\
$E\,\,\,$&$2\,$&$-1~$&$2~$&$0~$&$0~$\\
$T_1\,$&$3\,$&$0~$&$-1~$&$1~$&$-1~$\\
$T_2\,$&$3\,$&$0~$&$-1~$&$-1~$&$1~$
\end{tabular}
\end{center}
\caption{\label{table-3}\footnotesize{The character table of 
 the tetrahedral point group $T_d$.}}
\end{table}
We enumerate the elements of this discrete subgroup of $SO(3)\subset
SU(3)$ in the same traditional order in which they are listed in the
character table, namely identity, followed by the eight $C_3$
rotations, followed by the three $C_2$ rotations, followed by the six
$S_4$ improper rotations, and finally the six mirror reflections.

\noindent
Identity\,:
$$
\left(
\begin{array}{ccc}
1&0&0\\
0&1&0\\
0&0&1
\end{array}
\right),$$

\noindent
$C_3$ Rotations\,:
\begin{equation*}
\hspace{-1.5cm}\left(
\begin{array}{ccc}
0&1&0\\
0&0&1\\
1&0&0
\end{array}
\right),\;
\left(
\begin{array}{ccc}
0&1&0\\
0&0&-1\\
-1&0&0
\end{array}
\right),\;
\left(
\begin{array}{ccc}
0&-1&0\\
0&0&-1\\
1&0&0
\end{array}
\right),\;
\left(
\begin{array}{ccc}
0&-1&0\\
0&0&1\\
-1&0&0
\end{array}
\right),
\end{equation*}
\begin{equation*}
\hspace{-1.5cm}\left(
\begin{array}{ccc}
0&0&1\\
1&0&0\\
0&1&0
\end{array}
\right),
\left(
\begin{array}{ccc}
0&0&1\\
-1&0&0\\
0&-1&0
\end{array}
\right),
\left(
\begin{array}{ccc}
0&0&-1\\
-1&0&0\\
0&1&0
\end{array}
\right),
\left(
\begin{array}{ccc}
0&0&-1\\
1&0&0\\
0&-1&0
\end{array}
\right)
\end{equation*}

\noindent
$C_2$ Rotations\,:
$$\left(
\begin{array}{ccc}
1&0&0\\
0&-1&0\\
0&0&-1
\end{array}
\right),
\left(
\begin{array}{ccc}
-1&0&0\\
0&-1&0\\
0&0&1
\end{array}
\right),
\left(
\begin{array}{ccc}
-1&0&0\\
0&1&0\\
0&0&-1
\end{array}
\right)
$$

\noindent
Improper rotations $S_4$\,:
$$\left(
\begin{array}{ccc}
1&0&0\\
0&0&1\\
0&-1&0
\end{array}
\right),
\left(
\begin{array}{ccc}
1&0&0\\
0&0&-1\\
0&1&0
\end{array}
\right),
\left(
\begin{array}{ccc}
0&-1&0\\
1&0&0\\
0&0&1
\end{array}
\right),$$

$$
\left(
\begin{array}{ccc}
0&1&0\\
-1&0&0\\
0&0&1
\end{array}
\right),
\left(
\begin{array}{ccc}
0&0&-1\\
0&1&0\\
1&0&0
\end{array}
\right),
\left(
\begin{array}{ccc}
0&0&1\\
0&1&0\\
-1&0&0
\end{array}
\right)
$$

\noindent Mirror reflections $\sigma_d$\,:
$$\left(
\begin{array}{ccc}
-1&0&0\\
0&0&-1\\
0&-1&0
\end{array}
\right),
\left(
\begin{array}{ccc}
-1&0&0\\
0&0&1\\
0&1&0
\end{array}
\right),
\left(
\begin{array}{ccc}
0&-1&0\\
-1&0&0\\
0&0&-1
\end{array}
\right),$$
$$
\left(
\begin{array}{ccc}
0&1&0\\
1&0&0\\
0&0&-1
\end{array}
\right),
\left(
\begin{array}{ccc}
0&0&-1\\
0&-1&0\\
-1&0&0
\end{array}
\right),
\left(
\begin{array}{ccc}
0&0&1\\
0&-1&0\\
1&0&0
\end{array}
\right)
$$

We note that all these $24$ matrices are signed permutation matrices,
also called generalized permutation matrices. They all have unit
determinant, since we insisted that they be elements of $SO(3)$  and not $O(3)$. If
this requirement is relaxed, we get `another copy' of $24$ matrices,
obtained by multiplying the original $24$ by the negative of the unit
matrix, namely $-\mathds{1}$, adding to a total of $48$. Noting that
$-\mathds{1}$ commutes with all elements, we recall the manner in
which one goes, in the case of point groups, from the $24$ element
tetrahedral group $T_d$ to
the $48$ element octahedral group $O_h$  by simply adding a new
element `inversion' which commutes with all elements of $T_d$. Indeed,
the full set of all $48$ signed permutation matrices, with both
signatures allowed for the determinant, is isomorphic to $O_h$. But
this larger group is of no direct relevance for our present situation.

\noindent
{\bf Remark}\,:  At the risk of repetition we note that 
the matrices above which go under `improper rotations $S_4$' 
and `mirror reflections $\sigma_d$' are actually proper $SO(3)$  rotations. 
 It is  in anticipation of their action on certain three-sections of $\Omega_3$ that  
we have  
so named them.
 
We have seen that the eight-dimensional real linear span of the
$\lambda$-matrices decomposes,  under the conjugation action of our 
tetrahedral group of twenty four
$SO(3)$ matrices, 
into  a direct sum of three 
 (irreducible) orthogonal subspaces.  
In each 
subspace we will, therefore, expect to obtain a representation of
$T_d$. It is instructive to actually construct these representations.

To this end let us relabel the eight $\lambda$-matrices as ${\bm J} =
(J_1,\,J_2,\,J_3) \equiv (\lambda_2,\,\lambda_5,\,\lambda_7)$, ${\bm 
X} =
(X_1,\,X_2,\,X_3) \equiv (\lambda_1,\,\lambda_4,\,\lambda_6)$, and 
${\bm Q}  =
(Q_1,\,Q_2) = (\lambda_3,\,\lambda_8)$. Let us denote by ${R}$ a
generic element of our tetrahedral group 
$T_d\subset SO(3) \subset SU(3)$ of
signed permutation matrices, and let the $8\times 8$ matrix ${\rm Ad}\,(R)$
 be the image of ${R}$ in the adjoint representation. With ${\bm x} = 
(x_1,\,x_2,\,x_3),\,
{\bm y} =(y_1,\,y_2,\,y_3) \in \mathds{R}^3$ and 
${\bm z} = (z_1,\,z_2)\in
\mathds{R}^2$, an element in the (real) linear span of the
$\lambda$-matrices, which necessarily has the form 
${\bm x}\cdot {\bm X} + {\bm y}\cdot{\bm J} + {\bm z}\cdot{\bm Q}$
with uniquely determined ${\bm x},\,{\bm y},\,{\bm z}$, can be denoted 
$ {\bm n}={\bm x}\oplus {\bm y}\oplus{\bm z} \in \mathds{R}^8$. The point being 
made is that the 24 eight-dimensional matrices 
 representing action of   
 our group $T_d \subset SO(3)\subset SU(3)$ 
 on $\mathds{R}^8$ [the adjoint representation of SU(3)] {\em simultaneously assume a 
direct sum form}.  That is, for every $R\in T_d$  
\begin{eqnarray}
{\bm x}\cdot {\bm X} + {\bm y}\cdot{\bm J} + {\bm z}\cdot{\bm Q} 
&\to& R(\,{\bm x}\cdot {\bm X} + {\bm y}\cdot{\bm J} + {\bm z}\cdot{\bm 
Q}\,)R^T\nonumber\\ 
&=& {\bm x}'\cdot {\bm X} + {\bm y}'\cdot{\bm J} + {\bm z}'\cdot{\bm Q},
\end{eqnarray}
with
\begin{equation}
\eqalign{
{\bm y}'\cdot {\bm J} = R({\bm y}\cdot{\bm J})R^T = (T_1(R){\bm y})\cdot{\bm J},\\
{\bm x}'\cdot {\bm X} = R({\bm x}\cdot{\bm X})R^T = (T_2(R){\bm x})\cdot{\bm X},\\
{\bm z}'\cdot {\bm Q} = R({\bm z}\cdot{\bm Q})R^T = (E(R){\bm z})\cdot{\bm Q}.\label{**}}
\end{equation}
We may denote this action of $T_d$ on $\mathds{R} ^8$ or $\Omega_3$ in the 
following compact form: 
\begin{equation}
R\rightarrow {\rm Ad}\,(R) = T_1(R)\oplus T_2(R) \oplus E(R), 
\end{equation}
where the matrices $T_1(R),~T_2(R),~E(R)$ are defined through Eq.\,(\ref{**}).

We list below the matrices $T_1(R),~T_2(R),~E(R)$ corresponding to
${\rm Ad}\,(R)$ for each $R\in T_d$, and these are to be read as 
\begin{equation}
R\in T_d \subset SU(3) \to T_1(R)\oplus T_2(R) 
\oplus E(R) = {\rm Ad}(R)
\end{equation}

\noindent
Identity:\\
\begin{equation*}
\hspace{-2.5cm}\left(
\begin{array}{ccc}
1&0&0\\
0&1&0\\
0&0&1
\end{array}
\right)
\to
\left(
\begin{array}{ccc}
1&0&0\\
0&1&0\\
0&0&1
\end{array}\right)
\oplus
\left(
\begin{array}{ccc}
1&0&0\\
0&1&0\\
0&0&1\\
\end{array}\right)
\oplus
\left(
\begin{array}{cc}
1&0\\
0&1
\end{array}\right)
\end{equation*}

\noindent
$C_3$ Rotations:\\
\begin{equation*}
\hspace{-2.5cm}\left(
\begin{array}{ccc}
0&1&0\\
0&0&1\\
1&0&0
\end{array}
\right)
\to
\left(
\begin{array}{ccc}
\ 0&\ 0&\ 1\\
-1&\ 0&\  0\\
\ 0&-1&\ 0
\end{array}\right)
\oplus
\left(
\begin{array}{ccc}
0&0&1\\
1&0&0\\
0&1&0
\end{array}\right)
\oplus
\left(
\begin{array}{cc}
-1/2&\sqrt{3}/2\\
-\sqrt{3}/2&-1/2

\end{array}\right)
\end{equation*}
\begin{equation*}
\hspace{-2.5cm}\left(
\begin{array}{ccc}
\ 0&\ 1&\ 0\\
\ 0&\ 0&-1\\
-1&\ 0&\ 0
\end{array}
\right)
\to
\left(
\begin{array}{ccc}
\ 0&\ 0&-1\\
\ 1&\ 0&\ 0\\
\ 0&-1&\ 0
\end{array}\right)
\oplus
\left(
\begin{array}{ccc}
\ 0&\ 0&-1\\
-1&\ 0&\ 0\\
\ 0&\ 1&\ 0
\end{array}\right)
\oplus
\left(
\begin{array}{cc}
-1/2&\sqrt{3}/2\\
-\sqrt{3}/2&-1/2
\end{array}\right)
\end{equation*}
\begin{equation*}
\hspace{-2.5cm}\left(
\begin{array}{ccc}
\ 0&-1&\ 0\\
\ 0&\ 0&-1\\
\ 1&\ 0&\ 0
\end{array}
\right)
\to
\left(
\begin{array}{ccc}
\ 0&\ 0&\ 1\\
\ 1&\ 0&\ 0\\
\ 0&\ 1&\ 0
\end{array}\right)
\oplus
\left(
\begin{array}{ccc}
\ 0&\ 0&\ 1\\
-1&\ 0&\ 0\\
\ 0&-1&\ 0
\end{array}\right)
\oplus
\left(
\begin{array}{cc}
-1/2&\sqrt{3}/2\\
-\sqrt{3}/2&-1/2
\end{array}\right)
\end{equation*}
\begin{equation*}
\hspace{-2.5cm}\left(
\begin{array}{ccc}
\ 0&-1&\ 0\\
\ 0&\ 0&\ 1\\
-1&\ 0&\ 0
\end{array}
\right)
\to
\left(
\begin{array}{ccc}
\ 0&\ 0&-1\\
-1&\ 0&\ 0\\
\ 0&\ 1&\ 0
\end{array}\right)
\oplus
\left(
\begin{array}{ccc}
\ 0&\ 0&-1\\
\ 1&\ 0&\ 0\\
\ 0&-1&\ 0
\end{array}\right)
\oplus
\left(
\begin{array}{cc}
-1/2&\sqrt{3}/2\\
-\sqrt{3}/2&-1/2
\end{array}\right)
\end{equation*}
\begin{equation*}
\hspace{-2.5cm}\left(
\begin{array}{ccc}
0&0&1\\
1&0&0\\
0&1&0
\end{array}
\right)
\to
\left(
\begin{array}{ccc}
\ 0&-1&\ 0\\
\ 0&\ 0&-1\\
\ 1&\ 0&\ 0
\end{array}\right)
\oplus
\left(
\begin{array}{ccc}
0&1&0\\
0&0&1\\
1&0&0
\end{array}\right)
\oplus
\left(
\begin{array}{cc}
-1/2&-\sqrt{3}/2\\
\sqrt{3}/2&-1/2
\end{array}\right)
\end{equation*}
\begin{equation*}
\hspace{-2.5cm}\left(
\begin{array}{ccc}
\ 0&\ 0&\ 1\\
-1&\ 0&\ 0\\
\ 0&-1&\ 0
\end{array}
\right)
\to
\left(
\begin{array}{ccc}
0&1&0\\
0&0&1\\
1&0&0
\end{array}\right)
\oplus
\left(
\begin{array}{ccc}
\ 0&-1&\ 0\\
\ 0&\ 0&-1\\
\ 1&\ 0&\ 0
\end{array}\right)
\oplus
\left(
\begin{array}{cc}
-1/2&-\sqrt{3}/2\\
\sqrt{3}/2&-1/2
\end{array}\right)
\end{equation*}
\begin{equation*}
\hspace{-2.5cm}\left(
\begin{array}{ccc}
\ 0&\ 0&-1\\
-1&\ 0&\ 0\\
\ 0&\ 1&\ 0
\end{array}
\right)
\to
\left(
\begin{array}{ccc}
\ 0&-1&\ 0\\
\ 0&\ 0&\ 1\\
-1&\ 0&\ 0
\end{array}\right)
\oplus
\left(
\begin{array}{ccc}
\ 0&\ 1&\ 0\\
\ 0&\ 0&-1\\
-1&\ 0&\ 0
\end{array}\right)
\oplus
\left(
\begin{array}{cc}
-1/2&-\sqrt{3}/2\\
\sqrt{3}/2&-1/2
\end{array}\right)
\end{equation*}
\begin{equation*}
\hspace{-2.5cm}\left(
\begin{array}{ccc}
\ 0&\ 0&-1\\
\ 1&\ 0&\ 0\\
\ 0&-1&\ 0
\end{array}
\right)
\to
\left(
\begin{array}{ccc}
\ 0&\ 1&\ 0\\
\ 0&\ 0&-1\\
-1&\ 0&\ 0
\end{array}\right)
\oplus
\left(
\begin{array}{ccc}
\ 0&-1&\ 0\\
\ 0&\ 0&\ 1\\
-1&\ 0&\ 0
\end{array}\right)
\oplus
\left(
\begin{array}{cc}
-1/2&-\sqrt{3}/2\\
\sqrt{3}/2&-1/2
\end{array}\right)
\end{equation*}

\noindent
$C_2$ Rotations:\\
\begin{equation*}
\hspace{-2.5cm}\left(
\begin{array}{ccc}
\ 1&\ 0&\ 0\\
\ 0&-1&\ 0\\
\ 0&\ 0&-1
\end{array}
\right)
\to
\left(
\begin{array}{ccc}
-1&\ 0&\ 0\\
 0&-1&\ 0\\
 0&\ 0&\ 1
\end{array}\right)
\oplus
\left(
\begin{array}{ccc}
-1&\ 0&\ 0\\
\ 0&-1&\ 0\\
\ 0&\ 0&\ 1
\end{array}\right)
\oplus
\left(
\begin{array}{cc}
1&0\\
0&1
\end{array}\right)
\end{equation*}
\begin{equation*}
\hspace{-2.5cm}\left(
\begin{array}{ccc}
-1&\ 0&\ 0\\
\ 0&-1&\ 0\\
\ 0&\ 0&\ 1
\end{array}
\right)
\to
\left(
\begin{array}{ccc}
\ 1&\ 0&\ 0\\
\ 0&-1&\ 0\\
\ 0&\ 0&-1\\
\end{array}\right)
\oplus
\left(
\begin{array}{ccc}
\ 1&\ 0&\ 0\\
\ 0&-1&\ 0\\
\ 0&\ 0&-1
\end{array}\right)
\oplus
\left(
\begin{array}{cc}
1&0\\
0&1
\end{array}\right)
\end{equation*}
\begin{equation*}
\hspace{-2.5cm}\left(
\begin{array}{ccc}
-1&\ 0&\ 0\\
\ 0&\ 1&\ 0\\
\ 0&\ 0&-1
\end{array}
\right)
\to
\left(
\begin{array}{ccc}
-1&\ 0&\ 0\\
\ 0&\ 1&\ 0\\
\ 0&\ 0&-1
\end{array}\right)
\oplus
\left(
\begin{array}{ccc}
-1&\ 0&\ 0\\
\ 0&\ 1&\ 0\\
\ 0&\ 0&-1
\end{array}\right)
\oplus
\left(
\begin{array}{cc}
1&0\\
0&1
\end{array}\right)
\end{equation*}

\noindent
Improper rotations $S_4$\,:\\

\begin{equation*}
\hspace{-2.5cm}\left(
\begin{array}{rrr}
1&0&0\\
0&0&1\\
0&-1&0
\end{array}
\right)
\to
\left(
\begin{array}{ccc}
0&1&0\\
-1&0&0\\
0&0&1
\end{array}\right)
\oplus
\left(
\begin{array}{ccc}
0&1&0\\
-1&0&0\\
0&0&-1
\end{array}\right)
\oplus
\left(
\begin{array}{cc}
\frac{1}{2}&\frac{\sqrt{3}}{2}\\
\frac{\sqrt{3}}{2}&-\frac{1}{2}
\end{array}\right)
\end{equation*}
\begin{equation*}
\hspace{-2.5cm}\left(
\begin{array}{rrr}
1&0&0\\
0&0&-1\\
0&1&0
\end{array}
\right)
\to
\left(
\begin{array}{ccc}
0&-1&0\\
1&0&0\\
0&0&1
\end{array}\right)
\oplus
\left(
\begin{array}{ccc}
0&-1&0\\
1&0&0\\
0&0&-1
\end{array}\right)
\oplus
\left(
\begin{array}{cc}
\frac{1}{2}&\frac{\sqrt{3}}{2}\\
\frac{\sqrt{3}}{2}&-\frac{1}{2}
\end{array}\right)
\end{equation*}
\begin{equation*}
\hspace{-2.5cm}\left(
\begin{array}{rrr}
0&-1&0\\
1&0&0\\
0&0&1
\end{array}
\right)
\to
\left(
\begin{array}{ccc}
1&0&0\\
0&0&-1\\
0&1&0
\end{array}\right)
\oplus
\left(
\begin{array}{ccc}
-1&0&0\\
0&0&-1\\
0&1&0
\end{array}\right)
\oplus
\left(
\begin{array}{cc}
-1&0\\
0&1
\end{array}\right)
\end{equation*}
\begin{equation*}
\hspace{-2.5cm}\left(
\begin{array}{rrr}
0&1&0\\
-1&0&0\\
0&0&1
\end{array}
\right)
\to
\left(
\begin{array}{ccc}
1&0&0\\
0&0&1\\
0&-1&0
\end{array}\right)
\oplus
\left(
\begin{array}{ccc}
-1&0&0\\
0&0&1\\
0&-1&0
\end{array}\right)
\oplus
\left(
\begin{array}{cc}
-1&0\\
0&1
\end{array}\right)
\end{equation*}
\begin{equation*}
\hspace{-2.5cm}\left(
\begin{array}{rrr}
0&0&-1\\
0&1&0\\
1&0&0
\end{array}
\right)
\to
\left(
\begin{array}{ccc}
0&0&1\\
0&1&0\\
-1&0&0
\end{array}\right)
\oplus
\left(
\begin{array}{ccc}
0&0&-1\\
0&-1&0\\
1&0&0
\end{array}\right)
\oplus
\left(
\begin{array}{cc}
\frac{1}{2}&-\frac{\sqrt{3}}{2}\\
-\frac{\sqrt{3}}{2}&-\frac{1}{2}
\end{array}\right)
\end{equation*}
\begin{equation*}
\hspace{-2.5cm}\left(
\begin{array}{rrr}
0&0&1\\
0&1&0\\
-1&0&0
\end{array}
\right)
\to
\left(
\begin{array}{ccc}
0&0&-1\\
0&1&0\\
1&0&0
\end{array}\right)
\oplus
\left(
\begin{array}{ccc}
0&0&1\\
0&-1&0\\
-1&0&0
\end{array}\right)
\oplus
\left(
\begin{array}{cc}
\frac{1}{2}&-\frac{\sqrt{3}}{2}\\
-\frac{\sqrt{3}}{2}&-\frac{1}{2}
\end{array}\right)
\end{equation*}

\noindent
Mirror reflections $\sigma_d$:\\
\begin{equation*}
\hspace{-2.5cm}\left(
\begin{array}{rrr}
-1&0&0\\
0&0&-1\\
0&-1&0
\end{array}
\right)
\to
\left(
\begin{array}{ccc}
0&1&0\\
1&0&0\\
0&0&-1
\end{array}\right)
\oplus
\left(
\begin{array}{ccc}
0&1&0\\
1&0&0\\
0&0&1
\end{array}\right)
\oplus
\left(
\begin{array}{cc}
\frac{1}{2}&\frac{\sqrt{3}}{2}\\
\frac{\sqrt{3}}{2}&-\frac{1}{2}
\end{array}\right)
\end{equation*}

\begin{equation*}
\hspace{-2.5cm}\left(
\begin{array}{rrr}
-1&0&0\\
0&0&1\\
0&1&0
\end{array}
\right)
\to
\left(
\begin{array}{ccc}
0&-1&0\\
-1&0&0\\
0&0&-1
\end{array}\right)
\oplus
\left(
\begin{array}{ccc}
0&-1&0\\
-1&0&0\\
0&0&1
\end{array}\right)
\oplus
\left(
\begin{array}{cc}
\frac{1}{2}&\frac{\sqrt{3}}{2}\\
\frac{\sqrt{3}}{2}&-\frac{1}{2}
\end{array}\right)
\end{equation*}

\begin{equation*}
\hspace{-2.5cm}\left(
\begin{array}{rrr}
0&-1&0\\
-1&0&0\\
0&0&-1
\end{array}
\right)
\to
\left(
\begin{array}{ccc}
-1&0&0\\
0&0&1\\
0&1&0
\end{array}\right)
\oplus
\left(
\begin{array}{ccc}
1&0&0\\
0&0&1\\
0&1&0
\end{array}\right)
\oplus
\left(
\begin{array}{cc}
-1&0\\
0&1
\end{array}\right)
\end{equation*}
\begin{equation*}
\hspace{-2.5cm}\left(
\begin{array}{rrr}
0&1&0\\
1&0&0\\
0&0&-1
\end{array}
\right)
\to
\left(
\begin{array}{ccc}
-1&0&0\\
0&0&-1\\
0&-1&0
\end{array}\right)
\oplus
\left(
\begin{array}{ccc}
1&0&0\\
0&0&-1\\
0&-1&0
\end{array}\right)
\oplus
\left(
\begin{array}{cc}
-1&0\\
0&1
\end{array}\right)
\end{equation*}
\begin{equation*}
\hspace{-2.5cm}\left(
\begin{array}{rrr}
0&0&-1\\
0&-1&0\\
-1&0&0
\end{array}
\right)
\to
\left(
\begin{array}{ccc}
0&0&-1\\
0&-1&0\\
-1&0&0
\end{array}\right)
\oplus
\left(
\begin{array}{ccc}
0&0&1\\
0&1&0\\
1&0&0
\end{array}\right)
\oplus
\left(
\begin{array}{cc}
\frac{1}{2}&-\frac{\sqrt{3}}{2}\\
-\frac{\sqrt{3}}{2}&-\frac{1}{2}
\end{array}\right)
\end{equation*}
\begin{equation*}
\hspace{-2.5cm}\left(
\begin{array}{rrr}
0&0&1\\
0&-1&0\\
1&0&0
\end{array}
\right)
\to
\left(
\begin{array}{ccc}
0&0&1\\
0&-1&0\\
1&0&0
\end{array}\right)
\oplus
\left(
\begin{array}{ccc}

0&0&-1\\
0&1&0\\
-1&0&0
\end{array}\right)
\oplus
\left(
\begin{array}{cc}
\frac{1}{2}&-\frac{\sqrt{3}}{2}\\
-\frac{\sqrt{3}}{2}&-\frac{1}{2}
\end{array}\right)
\end{equation*}
Computing the characters (traces) of the matrices $T_1(R)$ and
comparing with the character table of $T_d$, we see that $T_1(R)$ is
actually nothing but the irreducible representation $T_1$ of
$T_d$. Similarly, $T_2(R)$ and $E(R)$ are verified to be the
irreducible representations $T_2$ and $E$ respectively. It is in
anticipation of this result that the matrices $T_1(R),~T_2(R),~E(R)$
were so named in the defining equations (\ref{**}). We have thus
 verified that the adjoint representation of $T_d \in SU(3)$, the
symmetry group of our obese-tetrahedron, is simply the direct sum of
three standard inequivalent irreducible representations of $T_d$.

We conclude this discussion of the symmetry of the obese-tetrahedron
with several comments. Firstly, we know that ${\rm Ad}\,(R)$, the image of
$R\in T_d\subset SO(3)\subset  SU(3)$  in the adjoint representation, 
is an element
of $SO(8)$, and 
hence should have positive determinant. This, however, does not
necessarily require that $T_1(R),~T_2(R)$ , and $E(R)$ should
individually have positive determinant. In particular, two of them can
have negative determinants. We note that $T_2(R)$ has positive
determinant for all `proper' rotations in $T_d$, but negative
 determinant for all the `improper' ones, namely the  $S_4$'s and 
$\sigma_d 
$'s, the signature in the latter
case being compensated by the signature of the determinant of
$E(R)$. Since $T_2$ is a vector representation, the above remark shows
that the triplet ${\bf X} = (\lambda_1,\,\lambda_4,\,\lambda_6)$ transforms 
as
a vector. On the other hand $T_1(R)$ has positive determinant for all
$R \in T_d$, including the improper ones. 
 That is, $T_1(R)$ supported by the triplet ${\bm J} = (\lambda_2,\,\lambda_5,\,\lambda_7)$ is a pseudo-vector
representation. 
This is consistent  with the fact that this
triplet actually comprises components of a pseudo-vector, the angular momentum.

Secondly, it may be noted that $R$ and $T_1(R)$
have the same character for every $R\in T_d$. 
 Thus it should be expected that the two
representations, $R$ and $T_1(R)$, are necessarily related by a change
of basis. The required change of basis is not hard to trace. Recall
that we listed the components of ${\bm J}$ as
$(\lambda_2,\,\lambda_5,\,\lambda_7)$, whereas the more appropriate
listing for components of angular momentum [generators of $SO(3)$] is
$(\lambda_7,\,-\lambda_5,\,\lambda_2)$. Thus the change of basis which will  
repair the listing is effected by
\begin{eqnarray}
R_0 &=& \left(\begin{array}{ccc}
0&0&1\\
0&-1&0\\
1&0&0
\end{array}\right)\in SO(3)\subset SU(3).
\end{eqnarray}
It can readily be verified that
\begin{equation}
R_0T_1(R)R_0^T = R,~\forall R\in T_d\subset SO(3)\subset SU(3).
\end{equation}

 Finally, the above analysis has at play in an interesting 
 manner the relationship among the full octahedral group 
 $O_h$  of order $48$, the $24$-element subgroup $O$ comprising the proper 
 rotations, and the $24$-element tetrahedral subgroup $T_d$. The  center of 
 $O_h$ is $Z_2 = (E,\,i)$ consisting of the identity and inversion, 
 and we have this important connection $O\times Z_2 =  O_h = T_d\times Z_2$ 
 The centre (of any group) acts trivially in the adjoint representation, 
 and it is for this reason that the full set of $48$ generalized permutation matrices 
constituting $O_h$ acts in  the adjoint representation  
 `\,as $O$\,'.%  on the linear span of $\lambda_2,\,\lambda_5,\,\lambda_7$ 
 % while it acts `\,as $T_d$\,' on the linear span of 
 % $\lambda_1,\,\lambda_4,\,\lambda_6$.   

%%
%%
\section{Final Remarks}
 We have presented in this paper a detailed analysis of the structure of the 
 generalized Bloch sphere $\Omega_3$, the state space of the qutrit. 
 We based our analysis on closed-form expressions for $\Omega_3$ and its boundary 
 $\partial \Omega_3$. That these expressions are nearly as economical in form 
 as the case of qubit was noted. Three special concentric spheres of radii $1$,  
 $1/\sqrt{3}$, and $1/2$ in $\mathds{R}^8$ were noted along with  
 their relevance to the structure of $\Omega_3$.

 We classified the $28$ standard two-sections of $\Omega_3$ into five unitary 
 equivalence classes of only four different geometrically inequivalent shapes. 
 That there are two different sets of circular two-sections that are 
 unitarily inequivalent but geometrically equivalent was noted, 
 a feature missed by earlier authors. The $56$ standard three-sections 
 were classified into ten unitary equivalence classes of 
 only seven different geometrically inequivalent shapes, correcting and completing 
an earlier classification by Mendas\,\cite{mendas}.
 The obese-tetrahedral three-section was examined in considerable detail, 
clarifying how its tetrahedral symmetry consisting of both proper and improper 
rotations arises from the $SO(3)\subset SU(3)$ group of proper rotations.  

From the point of view of the goal of this paper, it has provided complete
answers to the issues it set out to study. But from the larger perspective
of understanding the structure of quantum state space in general, what
this  work has accomplished is no more than just to scratch the very
surface,  and much remains to be done.
The present work has confined itself to standard
two and three-sections. It will be of interest to study the kind of  new
insights an analysis of generic two and three-sections could bring out.
Further, the present work is confined to qutrit. Generalization to higher
Hilbert space dimensions is necessary for relating issues of separability
and entanglement directly to the structural aspects of the geometry of
quantum state space.  We hope to return to these issues elsewhere.

%\SKG{The present work, although provides an interesting insight into the state space of qutrits, deals only with standard three-sections, which are not enough to understand the complex structure of the state space in its complete generality. It could be insightful to study arbitrary three-sections of a quantum state space of qutrits. Exploring the state space of higher dimensional quantum systems using the method of three-sections can also be a useful exercise.}

\vskip 0.3cm
\noindent
{\bf Acknowledgement}: The authors are grateful to Prof. Rajiah Simon
for permitting them to make liberal use, in Section~6, of his
unpublished Lecture Notes on {\em Point Groups}.

% \bibliographystyle{iopart-num}
% \bibliography{bibliography}

\providecommand{\newblock}{}

\end{document}